\documentclass[acmsmall]{acmart}
\usepackage[ruled,vlined]{algorithm2e}
\usepackage{xcolor}
\usepackage{multirow}

\AtBeginDocument{%
  \providecommand\BibTeX{{%
    \normalfont B\kern-0.5em{\scshape i\kern-0.25em b}\kern-0.8em\TeX}}}

\setcopyright{acmcopyright}
\copyrightyear{2018}
\acmYear{2018}
\acmDOI{10.1145/1122445.1122456}

\acmBooktitle{}
\acmPrice{15.00}
\acmISBN{978-1-4503-XXXX-X/18/06}

\begin{document}

\title{Anomaly Detection in Audio with Concept Drift using Adaptive Huffman Coding}

\author{Pratibha Kumari}
\affiliation{%
 \institution{Indian Institute of Technology, Ropar}
 \city{Ropar}
 \state{Punjab}
 \country{India}}
\email{2017csz0006@iitrpr.ac.in}

\author{Mukesh Saini}
\affiliation{%
 \institution{Indian Institute of Technology, Ropar}
 \city{Ropar}
 \state{Punjab}
 \country{India}}
\email{mukesh@iitrpr.ac.in}

\renewcommand{\shortauthors}{Kumari and Saini, et al.}

\begin{abstract}
\textcolor{black}{When detecting anomalies in audio, it can often be necessary to consider concept drift: the distribution of the data may drift over time because of dynamically changing environments, and anomalies may become normal as time elapses. We propose to use adaptive Huffman coding for anomaly detection in audio with concept drift. Compared with the existing method of adaptive Gaussian mixture modeling (AGMM), adaptive Huffman coding does not require a priori information about the clusters and can adjust the number of clusters dynamically depending on the amount of variation in the audio. To control the size of the Huffman tree, we propose to merge clusters that are close to each other instead of replacing rare clusters with new data. This reduces redundancy in the Huffman tree while ensuring that it never forgets past information. On a dataset of audio with concept drift which we have curated ourselves, our proposed method achieves higher area under the curve (AUC) compared with AGMM and fixed-length Huffman trees. The proposed approach is also time-efficient and can be easily extended to other types of time series data (e.g., video).}

\end{abstract}




\maketitle
\section{introduction}\label{introduction}
Initial automated surveillance systems were based on video sensors, which are quite popular even today. 
However, exclusively relying on video data can lead to unavoidable errors in monitoring. For example, the video analysis algorithms fail to work in the presence of adverse weather conditions, abrupt illumination changes, shadows, reflections, or when there is an obstacle on the line of sight ~\cite{crocco2016audio}. Moreover, a slight movement in the camera may cause background models to fail~\cite{stauffer1999adaptive}. 
Audio sensors are less sensitive to these errors. They provide a complementary aspect of the environment and facilitate a better scene understanding when used along with a video camera ~\cite{cristani2007audio,smeaton2005towards}. 
While there have been a large number of works on detecting suspicious activities (e.g., vehicle crashes, shouting, and gunshot detection) using audio analysis~\cite{crocco2016audio,foggia2015reliable,conte2012ensemble}, detecting anomalies is still a challenging task due to inherent concept drift.
\textcolor{black}{The phenomenon of concept drift refers to the change in the data distribution over time in dynamically changing and non-stationary environments~\cite{widmer1996learning}.
A typical example of concept drift is a change in noise level observed from day to night in an office. A noisy environment in the office appears normal during the working hours, whereas it becomes abnormal if observed in non-working hours. To accommodate concept drift, the model needs to be updated online, which is also referred to as adaptive learning.
Research related to learning with concept drift has been increasingly growing. \textcolor{black}{Many drift-aware adaptive learning algorithms have been developed  based on ARIMA ~\cite{bianco2001outlier}, sliding thresholds~\cite{widmer1996learning}, change point detection~\cite{basseville1993detection}, and exponential smoothing~\cite{szmit2012usage}. 
Recently, Hierarchical Temporal Memory~\cite{ahmad2017unsupervised} based model has emerged as a promising solution for learning under concept drift.
These approaches have been mainly explored for non-multimedia data such as text~\cite{ahmad2017unsupervised,costa2014concept}, power consumption data in smart cities~\cite{fenza2019drift,wang2020real}, web data~\cite{ma2018robust,lu2020data}, network data~\cite{ahmad2017unsupervised}, and financial data~\cite{ahmad2016real}}.
In spite of the popularity of this research topic, there is a lack of focus on concept drift handling for audio-visual data.}
There have been a few attempts to adapt GMM (Gaussian Mixture Model) for anomaly detection in audio \textcolor{black}~\cite{cristani2004line,cristani2007audio}. \textcolor{black}{However, the learning and forgetting strategy in an adaptive GMM (AGMM) approach reacts poorly to concept drift, which inherits a context of a longer time-span.} Moreover, such a modeling requires crucial prior knowledge about the number of clusters, their size, and their density. In addition, there is a need to calculate the inverse covariance matrix for Gaussians, which may even not exist in the case of multivariate data. 

Coding-cost based approaches do not need such prior knowledge; they are considered mostly parameter-free \cite{bohm2009coco}. 
In this article, we propose to use Huffman coding compression technique for audio anomaly detection. Huffman coding is a very simple yet effective technique for lossless data compression. It assigns more bits to the less probable data and less bit to frequent data. We have utilized this property to detect an anomaly, which is defined as a rare event. The existing works on the Huffman coding based anomaly detection method train a stationary model \cite{bohm2009coco}~\cite{callegari2009use}. To the best of our knowledge, we have not seen Huffman coding being applied for anomaly detection in time-series data with concept drift. A single model is trained and applied to the whole data \cite{bohm2009coco}~\cite{callegari2009use}. To cope up with concept drift in audio data, we utilize a modified version of the adaptive Huffman coding tree. 

There are numerous challenges in applying the Huffman coding on a data coupled with concept drift. The nodes in the tree can no longer be fixed. The Huffman tree should learn the environment in an online fashion. What leaf nodes represent also has to change with time. We learn an adaptive codebook for the leaf nodes using the audio bag-of-words approach \cite{lim2015robust}. Audio segments are converted to words, and a group of words is classified as an anomaly based on the coding cost. Along with the classification, the new data is also used to update the codebook as well as the tree structure.         
Traditionally, when a new sample does not match any of the Gaussians in the AGMM based approach, the weakest Gaussian is replaced with a new one corresponding to the new data \cite{cristani2004line,cristani2007audio}. We propose to use a node merging strategy instead of node replacement in the Huffman tree. After updating the tree parameters at the arrival of a new sample, relatively close tree nodes are merged. We experimentally found that the number of required nodes in the tree never explodes, no matter how long the data is. The node merging strategy also solves the problem of two data nodes drifting close to each other, which actually represent a single node. In this way, we never completely forget any past data. 

We also found that most of the existing public audio datasets have samples from stationary classes (also sometimes called fixed anomalies), e.g., glass breaking, gunshot, shouting, etc.~\cite{crocco2016audio,valenzise2007scream}. These are mostly short audio clips containing such events, mixed with miscellaneous background ~\cite{DCASE2017challenge}. Foggia et al. ~\cite{foggia2015audio} proposed a dataset that contains specific event samples useful for road surveillance. While these datasets are good for the evaluation of supervised classification techniques, they do not have concept drift. Audio clips are too short of having any drift in the data distribution. Therefore, we collect various challenging datasets that contain multiple natural anomalies (not staged). 
The dataset is available here\footnote{
   \href{https://drive.google.com/drive/folders/1T3JesUDB5Rlydl2wGBy6NEwrz9Aj2CLP?usp=sharing}{Dataset Link}}. Experiments on this dataset reveal that (1) the node merging scheme is more effective than node replacement, (2) the proposed method outperforms the previous method on adaptive anomaly modeling.

We brief our contributions as follows: 
\begin{itemize}
    \item We give a framework to apply Huffman coding to model concept drift in adaptive anomaly detection. The proposed method outperforms previous works and needs less prior knowledge.

    \item We propose a node-merging scheme for adapting the Huffman tree, which gives better results than the traditional node replacement scheme. 
    
    \item We propose a long duration audio dataset with concept drift for the evaluation of adaptive anomaly detection works. 
    
\end{itemize}

Rest of the paper is structured as follows. We do a literature survey in Section~\ref{sec:literature}. Then we present the proposed work in Section~\ref{sec:proposed}. The experimental results \& analysis are presented in Section~\ref{sec:exp}. We discuss limitations of the work in Section~\ref{sec:limitations}. Finally, Section~\ref{sec:conc} concludes the article.
\section{RELATED WORK}\label{sec:literature}
Audio analysis has been effectively used for detecting defined suspicious activities (e.g., vehicle crashes, shouting, and gunshot detection) both in indoor and outdoor scenarios~\cite{crocco2016audio,foggia2015reliable,conte2012ensemble}. Carletti et al.~\cite{carletti2013audio} separated gunshot, scream, broken glass, and background noise using multiple SVM classifiers, one for each type of anomalous event. Foggia et al.~\cite{foggia2015audio} used a similar approach for the task of car crashes and tire skidding detection. Rushe et al.~\cite{rushe2019anomaly} used a convolutional autoencoder-based semi-supervised approach to detect baby crying, broken glass, and gunshot. 
The above supervised methods are only effective in detecting stationary anomalies without any concept drift \cite{widmer1996learning}.

Real-world time-series data like weather data, audio streams, CCTV videos, etc., often have concept drift. 
A stationary anomaly model relies on the one-time learning of data distribution and hence performs poorly on the new unseen time series data. To detect an anomaly in the presence of concept drift, we need to adapt the model with time ~\cite{gama2014survey,zhao2017spatio,kratz2009anomaly,roshtkhari2013line}. 

Saurav et al.~\cite{saurav2018online} discussed concept drift issue for point anomaly detection where each sample had a semantic meaning. The authors proposed incremental learning of an RNN (Recurrent Neural Network) model by feeding the data present in a window of the given length.  
Fenza et al.~\cite{fenza2019drift} used LSTM (Long Short Term Memory) based drift aware model to detect anomalies in smart grid sensory data. They store the errors made in the last 24 hours to retrain the LSTM.  
The problem with the deep network-based approaches is that it takes a large number of samples and huge computing power to reflect any change of the distribution. In the case of audio, after mapping to the semantic space (feature calculation), we do not get as many examples.

AGMM is one of the successful models for anomaly detection in multimedia data. The baseline AGMM based approach~\cite{stauffer1999adaptive} was proposed for visual foreground-background detection. Each Gaussian in the mixture represents either a background or foreground sample. The anomalous samples have a lower likelihood of being generated from the mixture. The number of Gaussians is kept fixed (3-6). Over time, Gaussians representing rare events are replaced by newer anomalies; hence, the model only captures recent history. The model is later used to detect anomalies in audio data. Cristani et al.~\cite{cristani2004line,cristani2007audio} build feature-wise univariate AGMM. Each Gaussian in a mixture represents a possible aural word. Moncrieff et al.~\cite{moncrieff2007online} introduced an idea to keep recently deleted Gaussians in a fixed size cache to extend the model memory. Applying AGMM for audio anomaly detection requires a lot of prior information about the number of clusters, cluster size, density, etc. Furthermore, an AGMM framework requires the inverse covariance matrix for likelihood calculation, which may even not exist for multivariate data.
 
In this work, we propose to use Huffman coding as an alternative approach for anomaly detection in audio. The idea is that the coding cost is generally higher for an anomalous data segment \cite{bohm2009coco}. 

Huffman coding has been used in various anomaly detection applications like software obfuscation techniques~\cite{wu2010mimimorphism}, network traffic anomaly detection~\cite{sun2015anomaly}, etc.
Callegari et al.~\cite{callegari2009use} used standard Huffman coding for offline anomaly detection in TCP network traffic data. 
The authors found that it preformed better than clustering-based outlier detection approaches. 
B{\"o}hm et al.~\cite{bohm2009coco} used a probability density-based coding method for offline outlier detection in the data-mining field.
Uthayakumar et al. \cite{uthayakumar2017simple} also, employed Huffman coding (and some other compression methods) to detect anomalies in network data. All these methods perform one-time training of the Huffman tree. Moreover, to the best of our knowledge, the coding techniques have never been exploited for complex anomaly detection in multimedia data. 
In this work, we propose a framework to adapt the Huffman coding approach for audio data with concept drift. We propose a new idea of merging tree nodes, which gives better performance than the traditional strategy of replacing past data with new data.  
\begin{figure*}
\centerline{\includegraphics[width=1\linewidth]{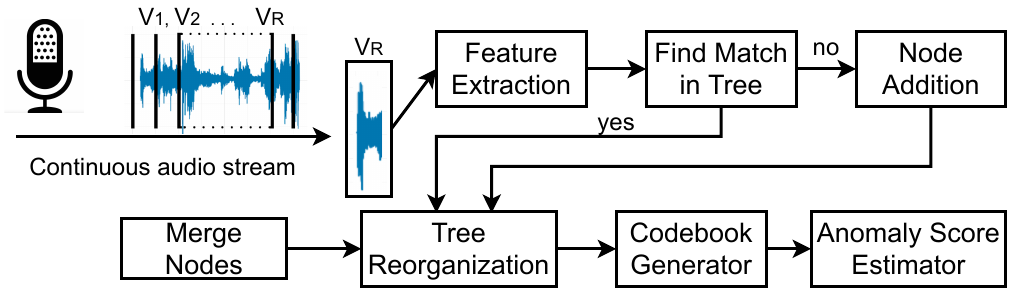}}
\caption{Coarse level architecture of the proposed framework.}
\label{fig:prop}
\end{figure*}
\section{PROPOSED WORK}\label{sec:proposed}
In this section, we first discuss the standard \textcolor{black}{dynamic} Huffman coding algorithm and then customize it to handle the inherent concept drift in audio anomaly detection task.
Figure~\ref{fig:prop} gives a high level overview of the proposed approach. 
We give a detailed description of each module of the framework in the below subsections.
\subsection{Dynamic Huffman Coding}\label{dynamicHuffCode}

Huffman coding algorithm is commonly used for lossless compression of data~\cite{huffman1952method}. A binary tree is built on probabilities of unique data samples. 
\textcolor{black}{These probabilities account for the frequency of occurrence of a particular sample and are treated as the weight of the corresponding node in the tree. Huffman algorithm is used to construct a binary tree with `n' external nodes and `n-1' internal nodes, where the external nodes store the weights $(w_{1},...w_{n})$. The tree produced by the Huffman algorithm has the minimum value of $\sum_{j}w_{j}l_{j}$ over all the binary trees, where $l_{j}$ is the level of the node with weight $w_{j}$. A binary tree that meets this condition is referred to as a correct Huffman tree, and the resulting codes are called optimal prefix codes.} The tree outputs a variable-length code (also known as minimal prefix code) for each data sample. Since the algorithm requires prior knowledge of data distribution, it is only suitable to encode offline data. 

The dynamic version of Huffman coding~\cite{knuth1985dynamic} allows encoding without prior knowledge of source distribution.
The tree evolves as it encounters new samples. Initially, there is a single node in the tree known as the root node. There is a concept of 0-node or NYT (not yet transmitted/traversed) label to a node. The root node is labelled as NYT in the beginning. The NYT node is broken into two child nodes when a new unique sample is encountered. The left child becomes the new NYT node, and the right child contains the data corresponding to the new unique sample. The left child is given a `0' weight, and the right child is assigned a weight of `1'. The weight of the parent node is the sum of the weights of the child nodes. For \textit{n} number of unique data samples, the tree posses a maximum of 2\textit{n}+1 (1 root, 1 NYT, \textit{n} data nodes, and (\textit{n}-1) internal nodes) number of nodes. Each node is given an unique index from 1 to 2\textit{n}+1. 
\textcolor{black}{If the root node is given an index of 1, then its right and left children are indexed 2 and 3. Thus, the indexing follows top to down and right to left order. Lower is the value of the node's index; higher is its rank in the Huffman tree.}

\textcolor{black}{The tree maintains a sibling property~\cite{knuth1985dynamic,vitter1987design} to ensure the correctness of the dynamic tree. The tree is said to possess a sibling property if each node has a sibling (except the root node), and they are numbered in order of non-decreasing weight with each node adjacent to its sibling. Also, the index of the parent node is lower than indexes of the children nodes.
Whenever there is an addition of a node or update in a node, the tree reorganizes itself by cascaded weight updates and swapping of the rule violating nodes. Thus it remains a \textcolor{black}{correct} Huffman tree for the data seen so far.}

\textcolor{black}{When a new sample matches any of the existing data nodes (let's name this node as $x$), its weight is to be increased by one. Before increasing the weight, the node's correct future position is determined. Positions of all the higher ranking nodes with the weight same as that of $x$ in the tree are considered. Then the node $x$ is swapped with the highest ranking node among the candidate nodes (let's name it as $\hat{y}$). \textcolor{black}{Note that $x$ and $\hat{y}$ are swapped along with sub-trees, if any.} After the swap, the weight of $x$ (in its new position now) is increased by one. Further, we go to the parent of $x$ and follow the same procedure to find its correct position in the tree. If we do not find any such node $\hat{y}$ then we simply increase $x's$ weight by one and go to its parent node. This way, we keep increasing the weight until we reach the root node.}

\textcolor{black}{Three types of swapping are possible while correcting the tree, namely, on the level ($\rightarrow$), up ($\uparrow$), and down ($\downarrow$). However, a $\downarrow$ swap is always followed by an immediate $\uparrow$ swap. Nodes at the different levels may be swapped.
The NYT node may also move up or down during the swap of nodes which contain the NYT node in their sub-trees. However, after the tree correction process finishes, the NYT node will always be at the lowest height (largest depth) in the tree since it has the lowest weight (0).
}

\textcolor{black}{In our case, weights are not integers but fractional numbers. In addition, on matching, weights are not increased by a fixed amount. The increment depends on the previous weight as well as the learning rate of the environment. Therefore, we modify the node swapping procedure while updating the tree (Section \ref{Tree Adaptation}). Further, we add a concept of merging nodes, which removes the constraint on the maximum number of nodes in the tree.}
\begin{algorithm}
    \nl \KwIn{Tree, $\vec{f}_{t}$, $ID_{NYT_{t-1}}$, $\theta_{cos}$, $\theta_{merge}$, Codebook}
    \KwOut{Codebook, $m_{t}$, score($\vec{f}_{t}$), $ID_{NYT_{t}}$}
    
    \nl \normalfont Search the tree to find a best match\\
    { match\_flag, $\hat{i}$, $S(\vec{\mu}^{\hat{i}}_t,\vec{f}_{t})$= Search\_Tree (Tree, $\vec{f}_{t}$, $ID_{NYT_{t-1}}$) using Equation~\ref{eq:search};
    }
    
    \nl \If{match\_flag==1 and $S(\vec{\mu}^{\hat{i}}_t,\vec{f}_{t})\geq \theta_{cos}$} 
    {
        update $w^{\hat{i}}_t$ \& $\vec{\mu}^{\hat{i}}_t$ using Equation~\ref{eq:weight},~\ref{eq:mu};\\
        recompute $\hat{i}$'s parent weight;\\
        $m_{t}\gets \hat{i}$;\\
        
    }
    \nl \Else
    {    
        replace the old NYT node with two children node such that left child is the new NYT node and right child is the data node corresponding to $\vec{f}_{t}$;\\
        
        $ID_{right child}\gets ID_{NYT_{t-1}}$+ 1\;\
        $ID_{NYT_{t}}\gets ID_{NYT_{t-1}}$ + 2;  \hspace{2em}// NYT node for future.\\
        $w_{right child}\gets w_{o}$;\\
        $w_{NYT_{t}}\gets 0$;  \\
        $w_{NYT_{t-1}}\gets w_{o}$; \\
        $\mu_{right child}\gets \vec{f}_{t}$;\\
        $m_{t} \gets ID_{right child}$;\\
    }

    \nl score($\vec{f}_{t}$)=$D(m_t)/  D(ID_{NYT_{t}})$; \hspace{2em}// D gives depth of a node using Codebook.\\

    \nl Collect all the \{$w^{i}_t$ , $\vec{\mu}^{i}_t$\} corresponding to leaf nodes in an array or list as:\\
    filtered\_w$\mu$ $\gets \{ w^{i}_t$ , $\vec{\mu}^{i}_t\}$;\\
    merge\_flag, merged\_w$\mu$=Recursive\_Merge($\theta_{merge}$, filtered\_w$\mu$);  \hspace{1em}// recursively finds two appropriate nodes to merge followed by deleting these two nodes and adding the output node to the list. \\
    \nl \If{merge\_flag==1 } 
    {
      Redraw a static Huffman tree with existing leaf nodes i.e., merged\_w$\mu$;\\
    }
    \nl \Else
    {    
    Reorganize\_Tree($m_{t}$, Tree) \hspace{2em}//corrects sibling property violations if any.\\
    }
    \nl Normalize $\{ (w^i_t) \mid \forall i \in\left \{ \mathcal{H} \right \} \}$ between 0-1 and update parents' weights accordingly;
    
    \nl Update the Codebook; \hspace{2em} // apply a tree traversal algorithm to visit each node and update their depth in the Codebook.\\
    
    \nl \textbf{procedure} Reorganize\_Tree(x, Tree)\\
    visitList$\gets$[x];\\
    \While{$x \geq 3$}
    {recompute weight of x if it is an internal node; \\
    $\hat{y} \gets \arg \max\limits_{y} (w_x-w_y) \forall y; ID_y<ID_x$;\\
    \If{($w_{x}-w_{\hat{y}}$) $>$ 0} 
    {     
      swap x and  $\hat{y}$ along with the subtrees (if any) and adjust the depth information;\\
      recompute parent's weight for nodes x \& $\hat{y}$;  \\
      append $\hat{y}$ in visitList if not in visitList;\\
    }
    z $\gets$ parent(x) if x is a right child otherwise sibling(x);\\
    append z in visitList if not in visitList and remove x from visitList;\\ 
    x$\gets$ lowest rank node from visitList;\\
    }
   recompute weight of node with ID=2 followed by node with ID=1; \\
   
    \caption{{\bf \textcolor{black}{Adapting the scene on arrival of a new audio frame}} \label{alg:Algorithm}}
\end{algorithm}
\subsection{Feature Extraction}
In a practical application, the audio captured by a surveillance device is mostly continuous in nature. The given audio is first divided into frames. Let $t$ represent the frame number, and $\Delta F$ be the duration of each frame. An event is represented by a set of continuous frames. Let $\mathcal{E}$ be the set of frame numbers that represent event $E$.    
We utilize the most frequently used audio features like MFCC (Mel-Frequency Cepstral Coefficients), energy, and ZCR (Zero Crossing Rate) to represent the audio signal. For a frame, we compute these features and concatenate them to get a column vector representation. Thus, we have a frame-wise feature vector representation of event $E$ as $\{ \vec{f}_{t}| t \in \mathcal{E}\}$.

\textcolor{black}{Feature normalization is an important step of data prepossessing in order to avoid any biases of the classifier towards numerically dominating features. However, for online settings, normalization or feature scaling is not straightforward as the actual mean and standard deviation of the data is not known a priori. Therefore we adopt a dynamic scaling technique~\cite{bollegala2017dynamic} where the approximated mean and standard deviation are dynamically computed for each of the features (energy, ZCR, and MFCC). These parameters are also updated on the arrival of each instance to accommodate the changes.}

\subsection{Tree Adaptation} \label{Tree Adaptation}
\textcolor{black}{We list all the steps of adapting the scene at the arrival of a new audio frame in Algorithm~\ref{alg:Algorithm}.}
After computing the low-level representation of $t^{th}$ frame ($\vec{f}_{t}$) and dynamic normalization, the existing adaptive Huffman tree is traversed from the top to bottom to find the best match (step 2). We use cosine similarity as a distance measure between the mean vector of $i^{th}$ data node ($\vec{\mu}^i_t$) and the current frame vector ($\vec{f}_{t}$). \textcolor{black}{The choice of cosine similarity measure is due to its bounded output, which makes it easier to keep a generic similarity threshold.
Each node in the tree is assigned a unique index (ID). Indexing of nodes starts from the root node with index `1', and it increases from top to down and right to left in the tree.}
Let $\hat{i}$ be the index of the best-matched node, which is computed as follows:
\begin{equation}
\hat{i}=\arg \max\limits_{i} \left [ S(\vec{\mu}^i_t,\vec{f}_{t}),\forall i \in\left \{ \mathcal{H} \right \} \right ]
\label{eq:search}
\end{equation}
where $\mathcal{H}$ is the set of indices of all the data nodes in the tree, and $S(\vec{\mu}^i_t,\vec{f}_{t})$ is a function that returns cosine similarity between mean vector of $i^{th}$ node $\vec{\mu}^i_t$ and $\vec{f}_{t}$. If $S(\vec{\mu}^{\hat{i}}_t,\vec{f}_{t})$ is greater than or equal to a threshold ($\theta_{cos}$), then this is called a hit. Hence, the hit node is a candidate node that has the largest similarity value. When we get a hit, the parameters (weight and mean) of the hit node are updated (step 3) as follows:
\begin{equation}
    {w}^{\hat{i}}_{t+1}=(1-\alpha)w^{\hat{i}}_t+ \alpha
\label{eq:weight}
\end{equation}
\begin{equation}
    \vec{\mu}^{\hat{i}}_{t+1}=(1-\gamma_{t})\vec{\mu}^{\hat{i}}_t+ \gamma_{t} \vec{f}_{t}
     \label{eq:mu}
\end{equation}
where ${w}^{\hat{i}}_t$ and $\vec{\mu}^{\hat{i}}_t$ represent the weight and the mean vector of $\hat{i}^{th}$ node at frame $t$ and ${w}^{\hat{i}}_{t+1}$ and $\vec{\mu}^{\hat{i}}_{t+1}$ at frame $t+1$; $\alpha$ is a learning parameter kept between 0 and 1; $\gamma_{t}$ is a update factor for mean vector corresponding to $\vec{f}_{t}$. It is inversely proportional to the distance between $\vec{\mu}^{\hat{i}}_t$ and $\vec{f}_{t}$. Cosine similarity varies between -1 to 1. Closer is the $S( \vec{\mu}^{\hat{i}}_t,\vec{f}_{t})$ to 1; higher is the update factor. The coefficient $\gamma_{t}$ is computed as follows:
\begin{equation}
     \gamma_{t}= \frac{S( \vec{\mu}^{\hat{i}}_t,\vec{f}_{t})-\theta_{cos}}{1-\theta_{cos}} \times   \left ( \gamma_{max}-\gamma_{min} \right ) +\gamma_{min}
     \label{eq:gammaupdate}
\end{equation}
\textcolor{black}{After updating the hit node, we recompute the weight of its parent node since the weight of one of its children nodes has changed. The weight of the parent is recomputed simply as the sum of the weights of the children nodes.} If we do not get a hit, then we break the current NYT node into the left child and right child (step 4). The node IDs for right and left children are assigned as $ID_{parent}$+1, $ID_{parent}$+2, respectively, where $ID_{parent}$ is the ID of node being split. Now the left child becomes the new NYT node for the future. The right child represents the new sample. \textcolor{black}{We also keep track of matched node ($m_{t}$) for $\vec{f}_{t}$ in order to compute the anomaly score at a later stage. $m_{t}$ is nothing but $\hat{i}^{th}$ node in case of a hit and the newly created right child in case of a miss.}
The initial mean vector for the right child node is set as the current frame vector. We give an initial weight ($w_{o}$) to this newly created data node. \textcolor{black}{The NYT node is given `0' weight, and thus the old NYT node gets a weight of $w_{o}$ (sum of children's weight).} 

\begin{figure}
\centerline{\includegraphics[width=1\linewidth]{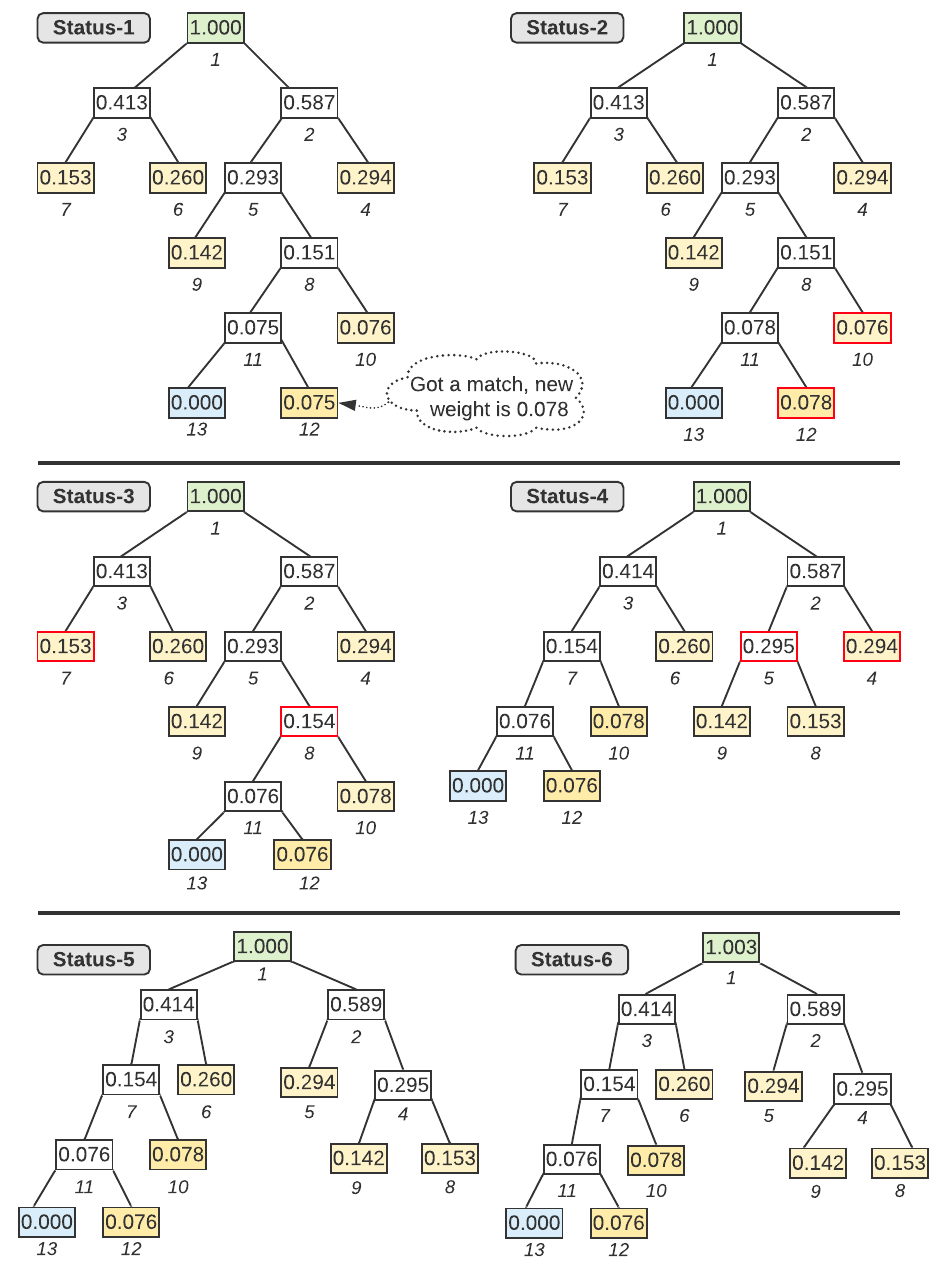}}
\caption{\textcolor{black}{An example of the reorganization process in the proposed algorithm. The colors of nodes: yellow, black, green, and white denotes a data node, the NYT node, the root node, and an internal node, respectively. A new audio frame matches with node-12 (Status-1). Adapting this new frame is shown via Status-2, 3, 4, 5, and 6. Red boxes in status-2, 3, and 4 highlight the nodes to be swapped.
}}

\label{fig:demoSync}
\end{figure}
After a hit/miss, we \textcolor{black}{reorganize} the tree in order to maintain the sibling property (step 8). \textcolor{black}{A node may violate the sibling property during the cascaded weight update process.
In that case, we first \textcolor{black}{swap the node violating this property with an appropriate node and }then continue the weight update process until we reach up to the root node. 
\textcolor{black}{We state the reorganization process in step 11 of Algorithm~\ref{alg:Algorithm}. We maintain a list, namely `visitList', of the node IDs that need to be visited in order to correct the tree. Initially, the list has the right child of the old NYT node in case of a miss and $\hat{i}^{th}$ node in case of a hit. Let's denote this node as $x$, and its weight as $w_{x}$. We take all the nodes with the ID lower (higher rank) than the ID of $x$ and weight lower than $w_{x}$. If we get any such nodes, then we swap $x$ with the node having the lowest weight among all the candidate nodes (let's denote this node as $\hat{y}$). The sub-trees are also swapped along with the nodes. The weight of the parent of node $x$ and $\hat{y}$ are also recomputed in the decreasing order of ID, i.e., the weight of higher ID is recomputed first then the lower ID. After this, we add the node $\hat{y}$ to the visitList. The successor node of $x$ (right sibling if $x$ is a left child else the parent node) is also added to the list. At this stage, the node $x$ is removed from the list. In case we do not find any such $\hat{y}$, we simply add the successor of $x$ and remove $x$ from the list. This way, we again take a new $x$ from the list as the node with the maximum index value (lower rank) and perform the reorganization. We keep repeating the process until we reach to right child of the root node. After reaching the right child of the root node, we simply recompute the weight of the root node. Note that the weight of the NYT node is not affected in the tree reorganization process, it remains at the lowest height in the tree. }}

\textcolor{black}{We demonstrate the tree reorganization process with an example tree in Figure~\ref{fig:demoSync}}. \textcolor{black}{Detailed steps are as follows: }
\begin{itemize}
    \item Status-1 in the figure shows a tree with 13 nodes. There are six unique data nodes (with IDs 4, 6, 7, 9, 10, and 12) in the tree. A new sample is matched with node-12 (for simplicity, we are denoting the node with ID 12 as node-12).
   
    \item The weight of node-12 is updated using Equation~\ref{eq:weight}, followed by recomputing the weight of its parent (node-11) as shown in Status-2. We find a suitable node $\hat{y}$ (node-10) for node $x$ (node-12) as explained in the above paragraph. 
    \textcolor{black}{Currently, the `visitList' consists of only node-12, i.e., `visitList'=[node-12].}
    
    \item We swap node-12 and node-10, followed by updating their parents' weights (status-3). Now node-10 and node-11 (parent of node-12) are added to the list, and node-12 is removed from the list. For the next round, we take the node with maximum index from the list, i.e., node-11, and find the corresponding $\hat{y}$. Here, we did not get any such $\hat{y}$, so we add its parent to the list, i.e., node-8, and delete node-11 from the list. In the further round, node-10 is taken out from the list. As there is no need to swap for node-10 as well, we push its parent to the list (in this case parent is already in the list, so we do not push it). \textcolor{black}{In this step, `visitList' is changed as follows: [node-10, node-11]->[node-10, node-8]->[node-8].}
    
    \item The next node taken out from the list is node-8. We get the corresponding $\hat{y}$ as node-7 and perform swapping as shown in `Status-4' in the figure. This time, node-7 and node-5 are added, and node-8 is removed. For the next round, node-7 is taken out. We do not find a $\hat{y}$ for node-7, so we push node-6 and remove node-7. Now the list has only node-6 and node-5, so node-6 is taken out. There is no need of swapping for node-6, so we push its parent, i.e., node-3, and remove node-6. \textcolor{black}{In this step, `visitList' is changed as follows: [node-8]->[node-7, node-5]->[node-5, node-6]->[node-5, node-3].}
    
    \item The next node taken out is node-5. We get node-4 as $\hat{y}$ for this node in the tree (Status-4). We swap these two nodes and recompute the parents' weight, followed by pushing node-2 and node-4 in the list (Status-5). Next, node-4 is taken out; there is no $\hat{y}$ for node-4. Node-4 is removed and node-2 is added (already in the list). \textcolor{black}{In this step, `visitList' is changed as follows: [node-5, node-3]->[node-3, node-2, node-4]->[node-3, node-2].}
    
    \item The next node taken out is node-3, and there is no need for swapping for this node. Now we are left with the right child of the root node in the list, so we finish reorganization procedure by just recomputing the weight of the root node (Status-6). \textcolor{black}{In this step, `visitList' is changed as follows: [node-3, node-2]->[node-2].}
    
\end{itemize}

The weight of a node may explode in the long run if we keep incrementing the weight. Therefore, we normalize weights of nodes between 0-1 at the end of each hit or miss (step 9). \textcolor{black}{Basically, we traverse the tree and take out all the weights of data nodes (all leaf nodes except the NYT node). Then we divide the weight of each data node by the summation of weights of all the data nodes. 
After this, the normalized weights of all the nodes are reflected by performing a postorder traversal on the tree.} \textcolor{black}{Thus, the NYT node has the lowest weight (0), and the root node gets the highest weight (1) after the weight normalization process.}
\subsection{Encoding Length Based Anomaly Score}
The next step of our algorithm is to generate the codebook. The codebook contains $\vec{\mu}^{i}$s (mean vectors of data nodes) and $D(i)$s (the distance of $i^{th}$ node from root node) (step 10). For anomaly detection, we do not need the actual codes but only the length of the code. The anomaly score is proportional to the code length, \textcolor{black}{i.e., the depth of the node. Along with the weight and the mean vector, we also store the depth information for each node. Whenever an NYT node is split, its children nodes are assigned a depth as $D(ID_{NYT_{t}})$+1. Also, at the time of a swap, the depth of the two nodes, as well as the attached sub-trees, is updated. The depth of nodes in the sub-tree can simply be updated by visiting all nodes using a pre-order traversal from the parent node (considering as a root node). Further, after processing the new sample, the $D(i)$s in the codebook can be updated by traversing the tree using any tree traversal algorithm.}
With the codebook and matched node ID ($m_{t}$), we can compute the anomaly score for the current audio frame. It is ratio of the depth of the matched node and the NYT node (step 5 of Algorithm~\ref{alg:Algorithm}). \textcolor{black}{There, the numerator term is the distance of the matched node from the root. $ID_{NYT_{t}}$ represents the ID of the NYT node in the current tree for $t^{th}$ frame. Thus, the denominator term ($D(ID_{NYT_{t}})$) is nothing but the depth of the tree for $t^{th}$ frame. Hence, the anomaly score of a frame is a relative distance of the matched node from the root node.
Further, the anomaly score ($\Omega$) for the event $E$ is computed by averaging the anomaly scores of all the frames of the event, i.e.,}
\begin{equation}
\Omega = \frac{1}{|\mathcal{E}|}\sum_{\forall t \in \mathcal{E}} 
\left ( \frac{D(m_t)}{ D(ID_{NYT_{t}})}   \right )
\label{eq:score}
\end{equation}

\subsection{Node Merging}
Contrary to the traditional methods, our algorithm does not keep an upper limit on the maximum number of nodes in the tree. The algorithm keeps adding the new unique frames in the tree on each miss. Thus, the codebook contains information about each unique event type seen so far, with reduced weight for events with a lower frequency of occurrence. Code length for a true anomalous frame is initially large as it does not have many neighbors. Other normal nodes have sufficient neighbors and thus remain in the upper part of the tree, leading to reduced code length. An abnormal frame that has seen sufficient neighbors may get upgraded to the upper level (toward the root node) of the tree through the \textcolor{black}{reorganization} process. Thus, the code length transition from large to small makes it a new normal. 

The nodes drift over time to accommodate the concept drift. The drifting nature may also bring two nodes near over time.

We introduce the concept of merging sufficiently close nodes. It is very likely that they represent the same type of event. At the arrival of each new frame, we check for the need to merge the nodes. We iteratively merge the two most similar data nodes if the distance between them is found to be below a merging threshold ($\theta_{merge}$ ) (step 6-7). \textcolor{black}{Thus, these two nodes can be at the same or different levels and not necessarily be siblings.}
\textcolor{black}{To perform the iterative merge, we take out all the data nodes by traversing the tree in a list or array data structure. We extract the two most similar nodes. If the similarity between these nodes is above $\theta_{merge}$ then we merge these nodes. We assign the weight of the output node as the sum of the weight of the two nodes. The mean vector for the output node is set as the point-wise average of the mean vectors of these nodes. \textcolor{black}{After this, we delete the two nodes from the list and add the merged node to the list. We keep merging until we do not find any such nodes to be merged. After this, we redraw the tree with existing nodes and weights similar to a fixed Huffman tree. Note that the redrawn tree is already a correct Huffman tree; hence we do not need to perform reorganization after a merge operation. For `n' data nodes, there will be a total of $n \choose 2$ unique data pairs. Hence, the time complexity to find such pair will be O($n^2$). For performing consecutive merge, where the array/list size shrinks by one element on each merge operation, the worst case time complexity will be O($n^3$).}}
The merging process lowers the memory requirement as well as reduces redundant false alarms due to redundant nodes. 
\begin{figure*}
\begin{minipage}[b]{.32\linewidth}
  \centering
 \includegraphics[width = 1\linewidth]{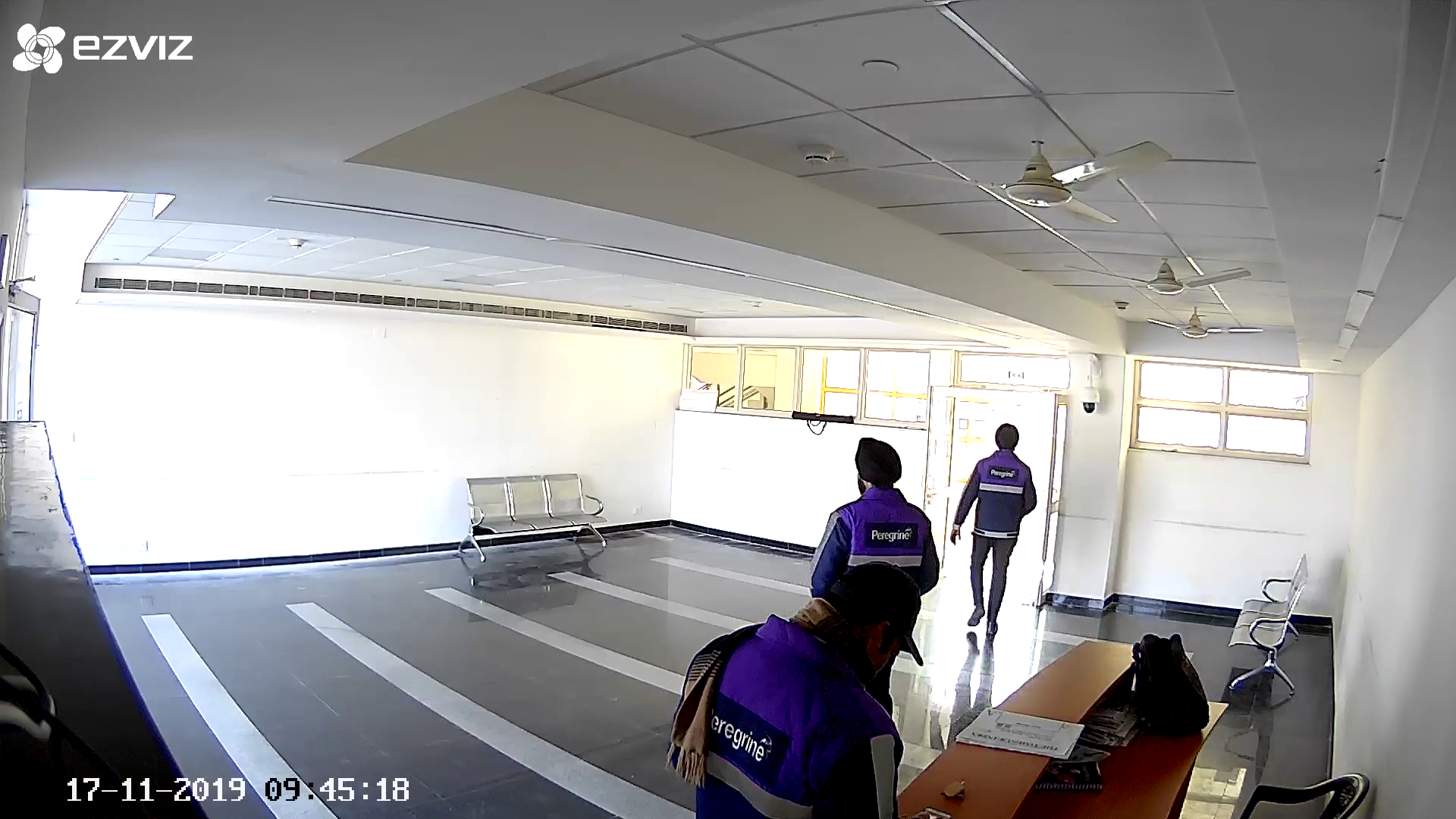}
  \centerline{(a) Reception Area (morning)}\medskip
\end{minipage}
\hfill
\begin{minipage}[b]{.32\linewidth}
  \centering
  \includegraphics[width = 1\linewidth]{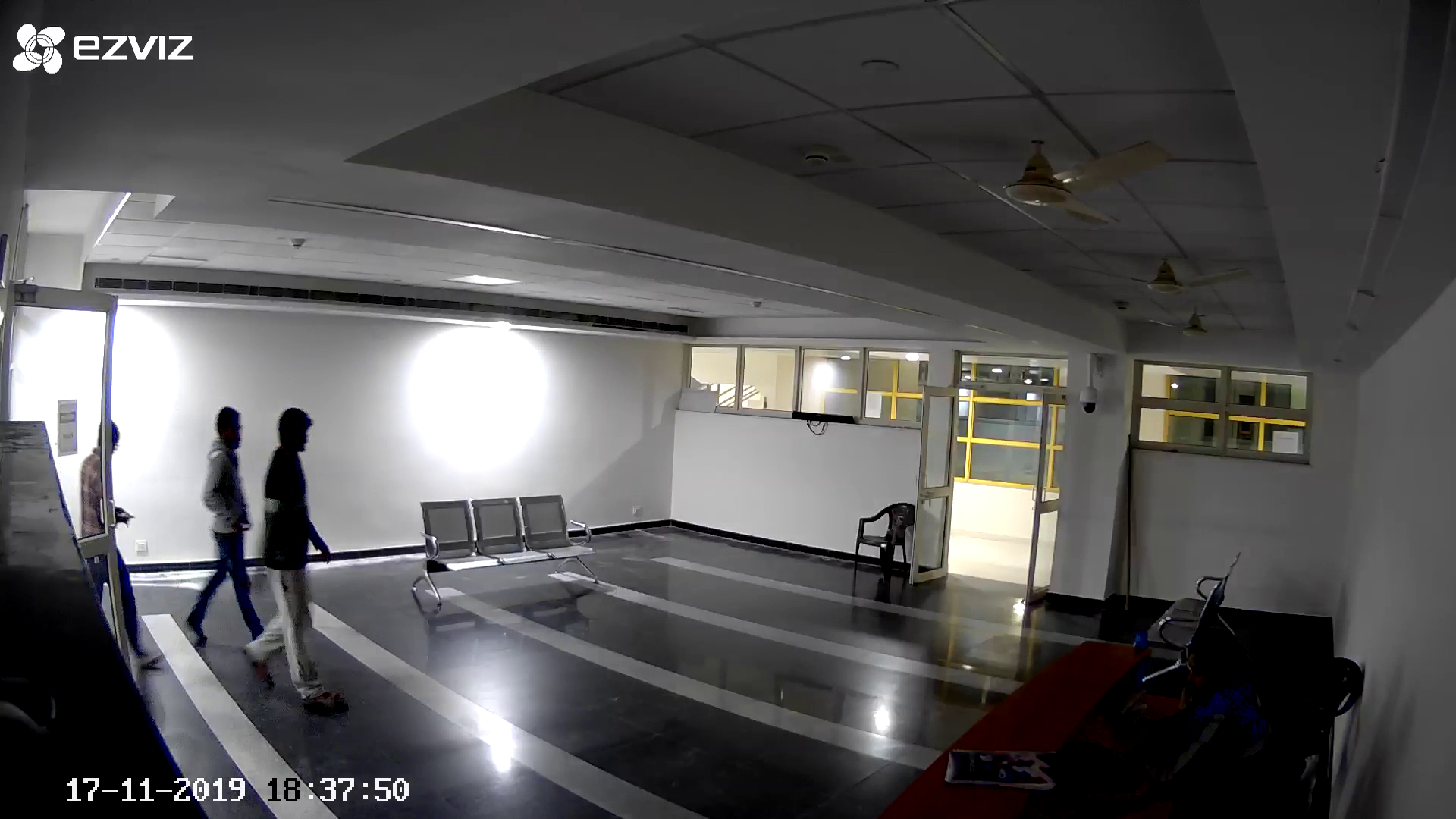}
  \centerline{(b) Reception Area (night)}\medskip
\end{minipage}
\hfill
\begin{minipage}[b]{.32\linewidth}
  \centering
  \includegraphics[width = 1\linewidth,height=7.1em]{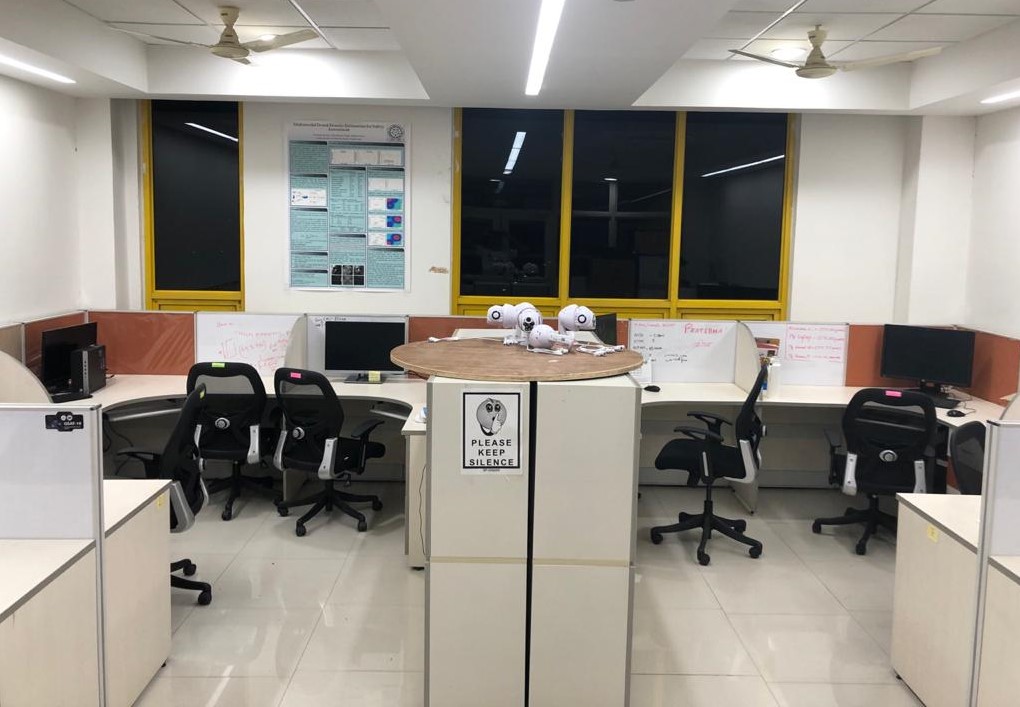}
  \centerline{(c) Lab-1 \& 2}\medskip
\end{minipage}
\begin{minipage}[b]{.32\linewidth}
  \centering
 \includegraphics[width = 1\linewidth]{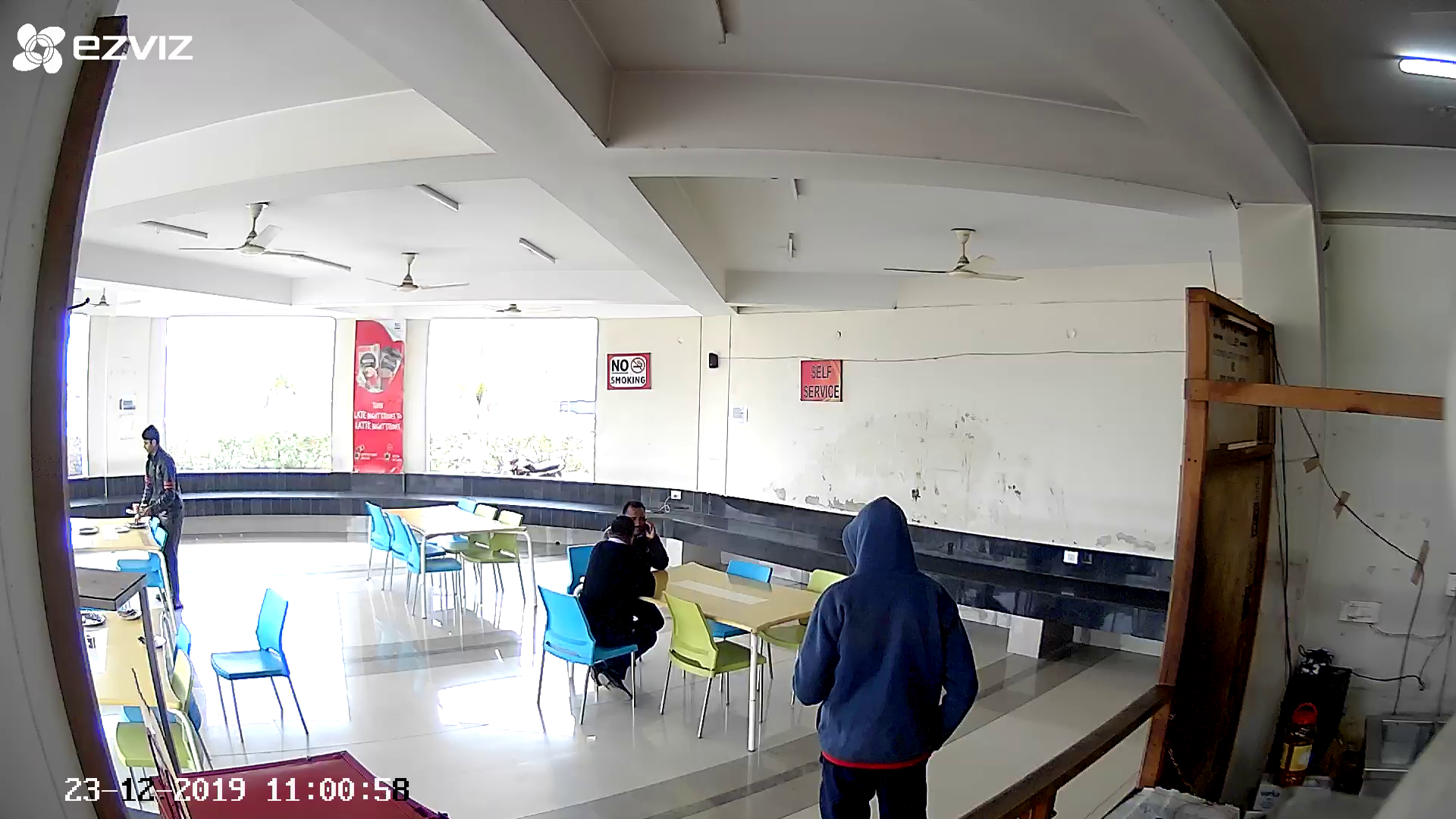}
  \centerline{(d) Outdoor Canteen (morning)}\medskip
\end{minipage}
\hfill
\begin{minipage}[b]{0.32\linewidth}
  \centering
  \includegraphics[width = 1\linewidth]{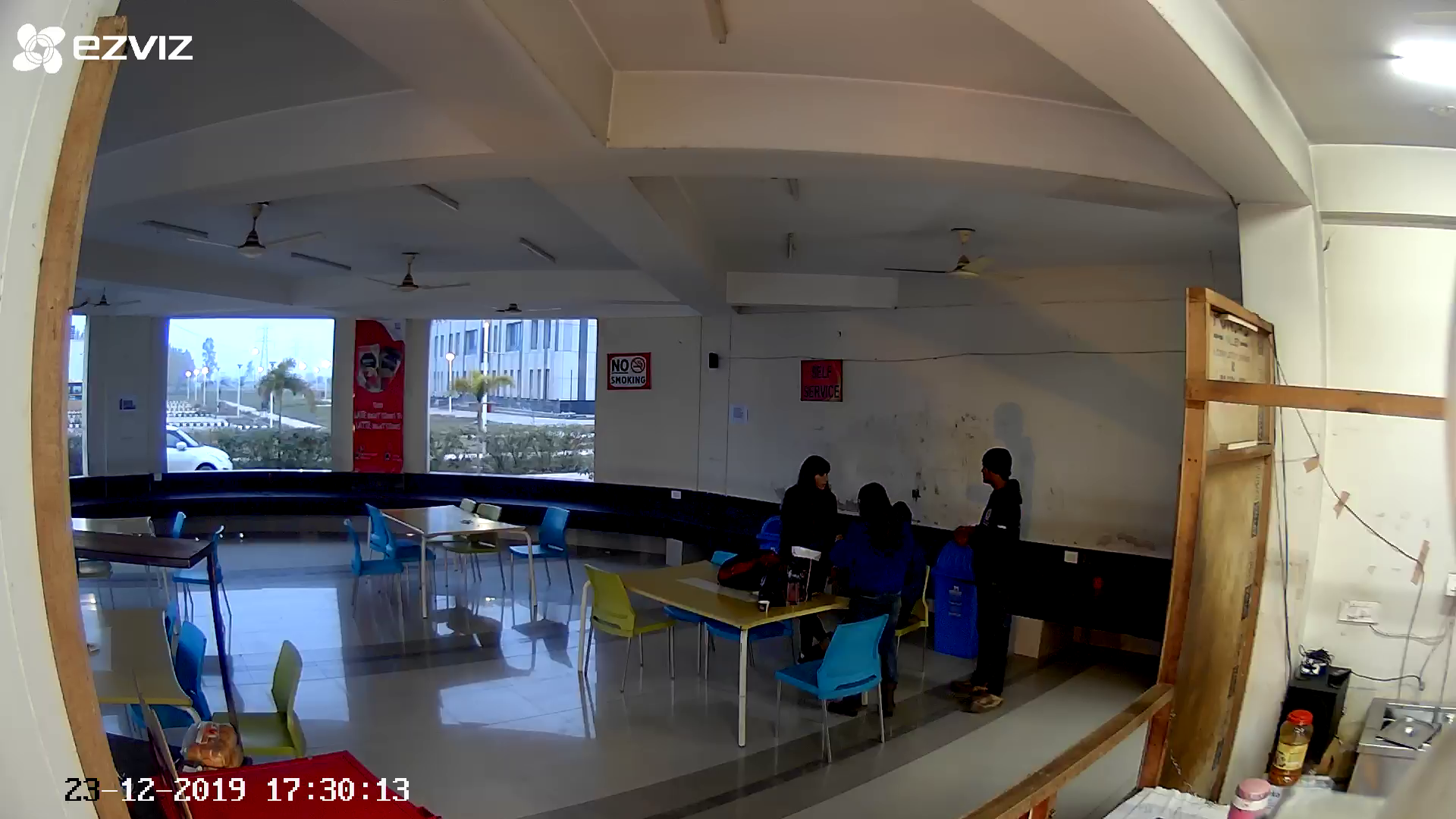}
  \centerline{(e) Outdoor Canteen (evening)}\medskip
\end{minipage}
\hfill
\begin{minipage}[b]{.32\linewidth}
  \centering
  \includegraphics[width = 1\linewidth]{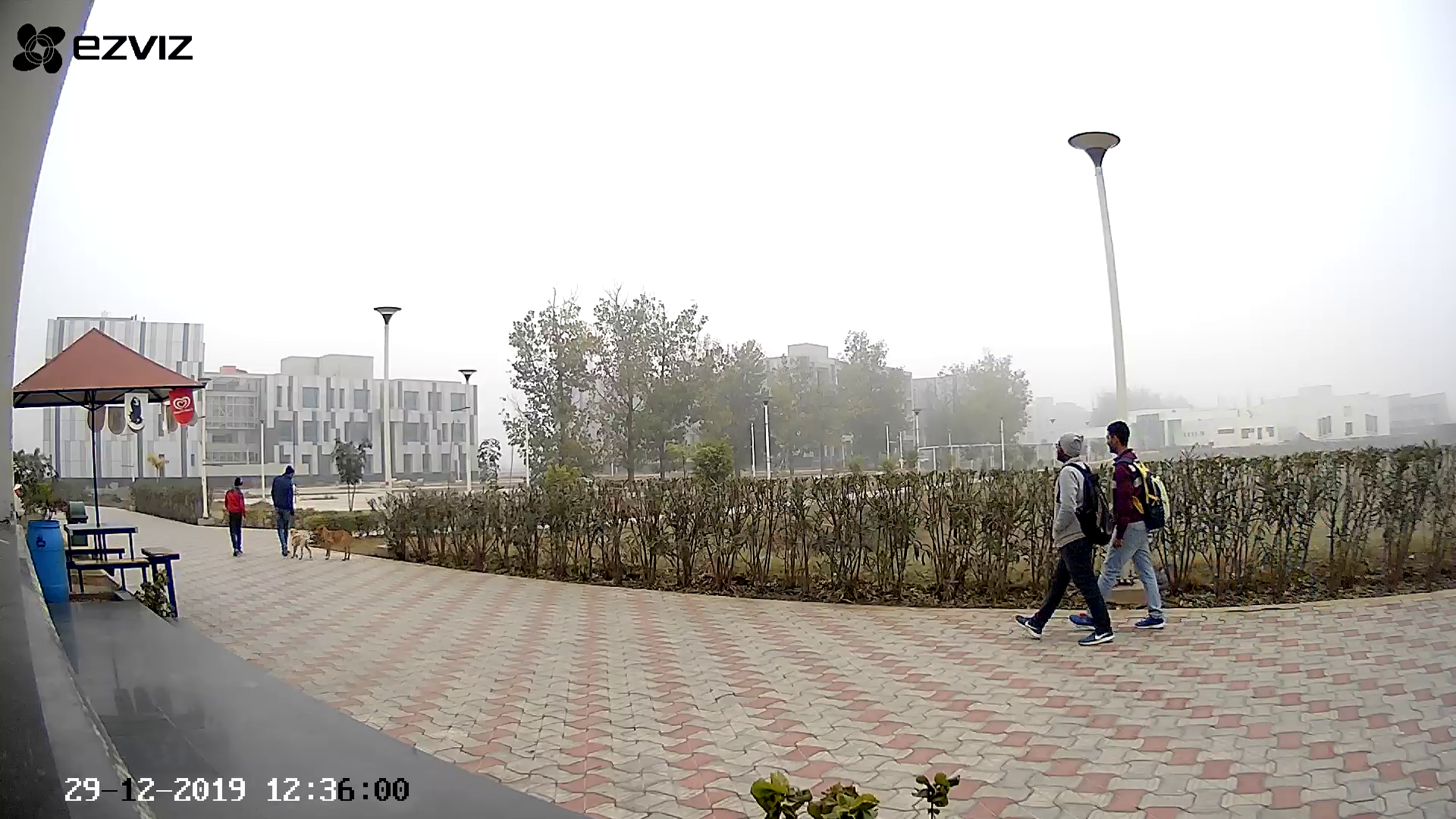}
  \centerline{(f) An Outdoor Area}\medskip
\end{minipage}
\caption{Representative images from the different data collection scenarios.}
\label{fig:sampleFrame}
\end{figure*}
\section{EXPERIMENTAL RESULTS}\label{sec:exp}
In this section, we discuss the datasets, experimental setup, and results. \textcolor{black}{We compare the proposed approach with a fixed Huffman coding-based approach and a baseline multivariate AGMM framework.}
\begin{table}
   \caption{\textcolor{black}{List of the datasets. `TV Series' and `12 Angry Men' datasets are taken from the web, and the rest others are self-collected.}}
  \label{tab:dataDescription}
  \begin{tabular}{ccccc}
    \toprule
    Dataset Name&Location/Description&Length&Purpose&Labeled\\
    \midrule
    Lab-1& a research lab &10 hours&No. of Nodes&No\\
    Reception Area& an institute&10 hours&No. of Nodes&No\\
    Outdoor Canteen-1& an open canteen&10 hours&No. of Nodes&No\\
    TV Series& Breaking Bad~\cite{breakingBad} &10 hours&No. of Nodes&No\\
   12 Angry Men~\cite{angryMen}& single room movie&1.5 hours &Performance&Yes\\
    Outdoor Canteen-2& an open canteen&\textcolor{black}{3 hours}&Performance&Yes\\
    An Outdoor Area& an institute&\textcolor{black}{3 hours}&Performance&Yes\\
    Lab-2& a research lab &1.33 hours&Adaptiveness&No\\
  \bottomrule
\end{tabular}
\end{table}
\begin{table}[]
\caption{\textcolor{black}{Percentage and example of anomalies in the datasets used for performance evaluation purposes.}}
  \label{tab:anomalyDescription}
\begin{tabular}{ccc}
\toprule
\multirow{2}{*}{Dataset} & \multirow{2}{*}{\begin{tabular}[c]{@{}c@{}}Anomaly \\ percentage\end{tabular}} & \multirow{2}{*}{Example anomalies} \\
 &  &  \\ \midrule
12 Angry Men & 2.27\% & Shout, laugh, door knock, water tap, raining, fan, etc. \\
Outdoor Canteen-2 & 1.57\% & \begin{tabular}[c]{@{}c@{}}Grinder sound, dragging chair, shout, vehicle passing, \\ music, sudden loud phone ringtone, fire siren, etc.\end{tabular} \\
An Outdoor Area & 2.6\% & \begin{tabular}[c]{@{}c@{}}Police siren, dog bark, floor cleaning machine, \\ laugh, music, bike, car, construction vehicle sound, etc.\end{tabular} \\ \bottomrule
\end{tabular}
\end{table}
\subsection{Datasets and Annotation Procedure} 
We utilized three datasets to show the efficacy of our work, namely `12 Angry Men' (1.5 hours), \textcolor{black}{`Outdoor Canteen-2' (3 hours), and `An Outdoor Area' (3 hours).} The `12 Angry Men'~\cite{angryMen} dataset is taken from the movie of the same name. The movie was shot in a single closed room. We collected the `Outdoor Canteen-2' and `An Outdoor Area' datasets by deploying a digital voice recorder, `SONY ICD-UX560F', with a sampling frequency of 22.05 kHz in two different outdoor environments. 
\textcolor{black}{We consider rare events as anomalies in our datasets. We want to mention that some of these rare events may not be a direct security threat (such as a gunshot), but they are still of interest to monitoring personnel; thus, all the low probable events are considered anomalies. Note that the probability of an event is not static due to concept drift, hence ground truth of an event might change over time.}
\textcolor{black}{`12 Angry Men' contains multiple natural anomalies like shouting, door knocking, sudden rain, laughing, etc. The sudden noise of grinder machine, fire siren, mixer noise, chair dragging, loud phone ringtone, etc., are some of the natural anomaly observed in the `Outdoor Canteen-2' dataset. The `An Outdoor Area' dataset is accompanied by more real-life outdoor anomalies, such as dog barking, police vehicle siren, sudden bike/car arrival, heavy construction vehicle passing by, etc. We give details of the anomaly duration and example anomalies for each of these datasets in Table~\ref{tab:anomalyDescription}.}

In addition to above mentioned datasets, we use four other datasets, each with a \textcolor{black}{duration} of 10 hours, to \textcolor{black}{study the maximum tree size (number of nodes in the tree).} 
These datasets are as follows: `TV Series' (Breaking Bad~\cite{breakingBad}), `Reception Area' (of an educational institute), `Outdoor Canteen-1', and `Lab-1'. The last three datasets are collected by us using the same digital voice recorder, and the first one is downloaded from the web. The choice of dataset covers a versatile range of outdoor scenarios. The `TV Series' dataset is chosen to show that the tree size doesn't burst even if we use multiple random audios mixed together. Apart from these datasets, we collected one more dataset, namely `Lab-2' of length `80' minutes (1.33 hours), \textcolor{black}{to assess the model's ability to remember events with long temporal context.} This dataset contains some intentionally created events. 
We played a song twice with an interval of 20 minutes \textcolor{black}{in order to see whether it was still considered an anomaly the second time it was played. The song was played for 98 and 17 seconds, respectively, during the two intervals.} All the above-described datasets are listed in Table~\ref{tab:dataDescription} with duration, recording location, and purpose. We also paste scene images via Figure~\ref{fig:sampleFrame} for better visualization of the datasets. Note that the camera is placed at the same location as the microphone to take the image.
\begin{table}
   \caption{\textcolor{black}{Cohen's kappa score between the annotator's pair on datasets used for performance evaluation.}}
  \label{tab:cohensKappa}
  \begin{tabular}{cccc}
    \toprule
    Annotator's Pair&12 Angry Men&Outdoor Canteen-2&An Outdoor Area\\
    \midrule
    $1^{st}$-$2^{nd}$&0.1537 &0.4460&0.2458\\
    $2^{nd}$-$3^{rd}$&0.1973 &0.6225&0.2445\\
    $3^{rd}$-$1^{st}$&0.5141 &0.6056&0.5999\\
  \bottomrule
\end{tabular}
\end{table}
\textcolor{black}{The datasets used for end result evaluation, namely `12 Angry Men', `Outdoor Canteen-2', and `An Outdoor Area' are annotated. Other datasets are not annotated because they are used to study model behavior not requiring a ROC/AUC. The three datasets have been manually annotated with the assistance of three security guards. Each guard is familiar with a maximum of one of the data collection scenarios. They are, however, unaware of the collected datasets as they were not on duty (they had a different shift) at the exact time of  recording. Also, they are unfamiliar with the movie dataset (`12 Angry Men'). They were shown a clip of the movie to gain some context for tagging. The security guards are from `Reception Area', `Outdoor Canteen', and `An Outdoor Area' sights (see Figure~\ref{fig:sampleFrame}). For annotation, they had to listen to the audio data with more focus. Additionally, individuals were asked to annotate independently, that is, without influencing other's opinions.} \textcolor{black}{They were instructed to label interesting or attention-catching phenomena as an anomaly.}

\textcolor{black}{They are given the three audio datasets to listen to and note down the start and end times of the anomalous segments. Both `An Outdoor Area' and `Outdoor Canteen' datasets have six half-hour chunks; `12 Angry Men' has a single chunk of 1.5 hours. They are not provided with any fixed set of events; rather, they can use their own judgment to define a situation and decide whether to consider it as novel (anomaly) or not.} \textcolor{black}{If an anomaly is observed for enough time, its all future occurrences are treated as normal. Again, the drift from abnormal to normal is subjective; hence we opted for a majority voting strategy for the final annotation of each dataset.}
\textcolor{black}{
\subsubsection*{Labeling and Inter-rater agreement analysis: }Let $\{T^k_1-T^k_2, T^k_3-T^k_4, T^k_5-T^k_6, ... \}$ be the the annotation list ($A^k$) of non-overlapping time-segments by $k^{th}$ annotator. 
The ground truth for all these time-segments by the security operator is `1', i.e., anomalous. 
With this information, we generate second-wise labels ($\vec{L}_k$) for each annotation list $A^k$, i.e., we label each second of the dataset as 1 (anomalous) or 0 (normal). 
If a particular second falls in any segment of $A^k$, we assign it a `1' else a `0'. 
In Table~\ref{tab:cohensKappa} we report Cohen's kappa~\cite{cohen1960coefficient}, an inter-annotator agreement measure for each of these datasets. The table shows that the Kappa values are in the range of 0.15-0.62. The mapping of Cohen Kappa scores to the level of agreement~\cite{landis1977measurement} suggests that the agreement between different annotator's varies from \textit{slight agreement} to \textit{substantial agreement} for different datasets. The agreement is less on the `12 Angry Men' dataset compared to other datasets. 
Overall we see a moderate level of agreement, which might even be considered good for such a complex and subjective task of scene level anomaly annotation.}

\textcolor{black}{Once we have individual annotation vectors of a dataset ($\vec{L}_1$, $\vec{L}_2$, and $\vec{L}_3$), the aggregated second-wise annotation ($\vec{L}_{final}$) is computed using majority voting on these three vectors. For each second, the extracted majority element serves as the final annotation, i.e., if two or more annotators are saying abnormal, then it is abnormal, else normal. Further, with the help of second-wise annotations, we compute event-wise annotation. From $\vec{L}_{final}$, the ground truth for each $\Delta F$ second event (with some overlap) is computed. Basically, we take $\Delta F$-seconds values consisting of 0's and 1's from the $\vec{L}_{final}$ and compute the ground truth as the majority element. In case of a tie (equal number of 0's and 1's), the event is assigned label `0'. We term the event-wise ground truth vector as $\vec{E}_{final}$.}
\textcolor{black}{We have used (2/3) volume of each of these datasets for hyper-parameter selection and the rest for performance evaluation.}
\subsection{\textcolor{black}{Experimental Setup}}
We test the approach on an event size of 4 seconds \textcolor{black}{(extracted with 50\% overlap)}. However, the approach can be used for any length of the event because the tree is built for the fixed frame size ($\Delta F$), not the event ($E$). We fix the length of the frame as 0.5 seconds. Thus, there are a total of 8 frames in an event of 4 seconds. \textcolor{black}{Algorithm~\ref{alg:Algorithm} runs on each frame of an event and produces frame-wise anomaly scores. Anomaly scores of all the 8 frames of the event are aggregated and fed to Equation~\ref{eq:score}. 
Thus, we get event-wise anomaly scores. These predicted anomaly scores lie between 0 and 1. Higher is the anomaly score; more is the novelty of the event. These event-wise predicted anomaly scores and the corresponding binary ground truths ($\vec{E}_{final}$) are used for performance evaluation.} 
\textcolor{black}{For performance evaluation with a skewed class distribution like anomaly detection, ROC (receiver operating characteristic curve) and AUC (area under the ROC curve) measures have been generally used in literature. 
ROC curve illustrates the performance of a binary classifier as the discrimination threshold for the two classes is varied. It is a plot between the true positive rate (tpr) and the false positive rate (fpr) at multiple threshold values. For the Huffman tree-based approaches (it is a soft-type classifier which produces raw anomaly score rather than 0/1 hard label), ROC points were generated using the algorithm proposed in~\cite{fawcett2006introduction}.
The AUC score is computed as the area under the ROC curve by adding the successive areas of trapezoids~\cite{bradley1997use}. It  gives a single scalar value for each ROC curve, which is used to compare various algorithms. 
}

We keep $w_{o}$=1e-3, $\gamma_{max}$=0.5, and $\gamma_{min}$=0 throughout the experiments. The hyper-parameters ($\left \{ \theta_{merge}, \theta_{cos}, \alpha \right \}$) need to be tuned for each surveillance environment, which is a common practice in adaptive anomaly detection frameworks~\cite{stauffer1999adaptive,cristani2007audio}. 
\subsection{\textcolor{black}{Hyper-parameter Tuning}}
\textcolor{black}{To achieve the best performance, the hyper-parameters need to be tuned according to the targeted surveillance environment. We can shrink the search space with prior expert knowledge. For example, if the scene is about a static indoor environment, the similarity threshold should be kept lower compared to a dynamic environment. 
Also, for a frequently changing environment, the weight update rate should be high; otherwise, it will not be able to adapt to the speedy variations in the environment.
The experimental results for selecting the optimum hyper-parameters on the `Outdoor Canteen-2' dataset are given in Table~\ref{tab:hyperparm}. There are 5394 events of 4 seconds in the `Outdoor Canteen-2' dataset. Hence, 2/3 fraction, i.e., 3,596 events, were used for hyper-parameter selection. As a new audio frame arrives, it is matched against the existing tree, an anomaly score is assigned accordingly, and finally, the tree is updated as well. Likewise, the anomaly score of all the audio frames in one event is collected to predict the anomaly score of the event using Equation~\ref{eq:score}. Further, we accumulate the anomaly score of all the events in the train data and compute the AUC value.
We varied the model hyper-parameters $\alpha$, $\theta_{cos}$, and $\theta_{merge}$ in range \{1e-05-1e-01\}, \{0.88-0.95\}, and \{0.96-0.99\} respectively and reported the corresponding AUC measures in Table~\ref{tab:hyperparm}.
If we see average AUC behavior per $\alpha$ value for a fixed value of $\theta_{cos}$, we observe that model is overall performing better with an $\alpha$=1e-04 (with some exceptions). Likewise, when we see average AUC for each $\theta_{cos}$ for a fixed value of $\alpha$, we find that a similarity threshold ($\theta_{cos}$) of 0.93 or 0.94 is giving better results in this scenario. 
We conclude that the model attains the best performance on $\alpha$=1e-04, $\theta_{merge}$=0.97, and $\theta_{cos}$=0.93. Using the selected values of hyper-parameters, we report AUC and ROC on the test datasets and found that performance does not degrade. Hence, we say that the parameter for an environment needs to be tuned once, and later, it can be used for that particular environment. The hyper-parameters for other scenarios are also tuned in the similar way. The model attains the best performance for the other scenarios at the following ($\theta_{cos}$, $\alpha$, and $\theta_{merge}$) values: `12 Angry Men' (0.88, 1e-04, and 0.98), `An Outdoor Area' (0.94, 1e-03, and 0.98).
}
\begin{table}[]
\centering
\caption{\textcolor{black}{AUC performance on various choices of $\alpha$, $\theta_{cos}$, and $\theta_{merge}$ for the `Outdoor Canteen-2' dataset.}}
\label{tab:hyperparm}
\begin{tabular}{ccccccccc}
\toprule
\multicolumn{1}{c|}{$\alpha$} & \multicolumn{1}{c|}{\begin{tabular}[c]{@{}c@{}}$\theta_{cos} \rightarrow$\\ $\theta_{merge} \downarrow$\end{tabular}} & 0.88 & 0.90 & 0.91 & 0.92 & 0.93 & 0.94 & 0.95  \\ \hline

\multicolumn{1}{c|}{\multirow{4}{*}{$10^{-5}$}} & \multicolumn{1}{c|}{0.96} & 65.20 & 65.05 & 67.08&70.42  &78.24  &74.78 &75.37  \\
\multicolumn{1}{c|}{} & \multicolumn{1}{c|}{0.97} &  62.93&68.87 &74.32  &72.41  &80.00 &76.31  &73.17\\
\multicolumn{1}{c|}{} & \multicolumn{1}{c|}{0.98} & 57.33 &67.24 &73.87 & 78.55 &77.18  &71.19 &73.14 \\
\multicolumn{1}{c|}{} & \multicolumn{1}{c|}{0.99} & 64.83 &53.97 &63.77 & 78.19 & 72.29 &70.31 &77.89   \\ \hline
\multicolumn{2}{c}{average AUC} & 62.57  &63.78 &69.76  &74.89  &\textbf{76.93} &73.15  &74.89 \\ \hline

\multicolumn{1}{c|}{\multirow{4}{*}{$10^{-4}$}} & \multicolumn{1}{c|}{0.96} &67.33  &68.46 &68.02  &70.57&76.53 & 75.12 &72.17 \\
\multicolumn{1}{c|}{} & \multicolumn{1}{c|}{0.97} &65.19  &67.63 &70.80  &72.15  &79.18 &74.64  &73.18\\
\multicolumn{1}{c|}{} & \multicolumn{1}{c|}{0.98} &58.94  & 64.21&  78.89& 76.12 &78.29 &75.49  &75.13\\
\multicolumn{1}{c|}{} & \multicolumn{1}{c|}{0.99} &69.25  &62.94 &69.70  &75.76  &74.51 &74.91  &73.75\\ \hline
\multicolumn{2}{c}{average AUC} &65.18  &65.81 &71.85  &73.65  &\textbf{77.13} &75.04  &73.56 \\ \hline

\multicolumn{1}{c|}{\multirow{4}{*}{$10^{-3}$}} & \multicolumn{1}{c|}{0.96} &67.23 &67.97 &69.31  &68.35 &73.08 &72.45  &68.71\\
\multicolumn{1}{c|}{} & \multicolumn{1}{c|}{0.97} &66.64  &69.05 & 69.51 &73.24  &75.22 &73.57  &72.98\\
\multicolumn{1}{c|}{} & \multicolumn{1}{c|}{0.98} & 63.56 &68.37 &74.65  &71.68  &73.67 &75.71  & 70.29\\
\multicolumn{1}{c|}{} & \multicolumn{1}{c|}{0.99} &71.25  &66.09 & 73.15 &67.22  &67.21 &68.63  &69.48\\ \hline
\multicolumn{2}{c}{average AUC} &67.17 & 67.87& 71.66 &70.12  &72.30 &\textbf{72.59}  &70.37 \\ \hline

\multicolumn{1}{c|}{\multirow{4}{*}{$10^{-2}$}} & \multicolumn{1}{c|}{0.96} &68.52 &68.08 &68.81  &68.87  &70.08 &70.63  &68.21 \\
\multicolumn{1}{c|}{} & \multicolumn{1}{c|}{0.97} & 68.93 & 70.40&68.69  &71.22  &69.94 & 69.92 &69.38\\
\multicolumn{1}{c|}{} & \multicolumn{1}{c|}{0.98} & 64.65 &69.36 &68.74  &71.05  &71.46 &73.76  &65.78\\
\multicolumn{1}{c|}{} & \multicolumn{1}{c|}{0.99} & 64.34 &67.75 & 68.39 &67.65  &67.83 &67.26  &70.02 \\ \hline
\multicolumn{2}{c}{average AUC} &66.61 &68.90 &68.66  &69.70  &69.83 &\textbf{70.39}  &68.35 \\ \bottomrule
\end{tabular}
\end{table}
\subsection{Maximum Number of Nodes in The Tree}
\begin{figure}
\centerline{\includegraphics[scale=0.55]{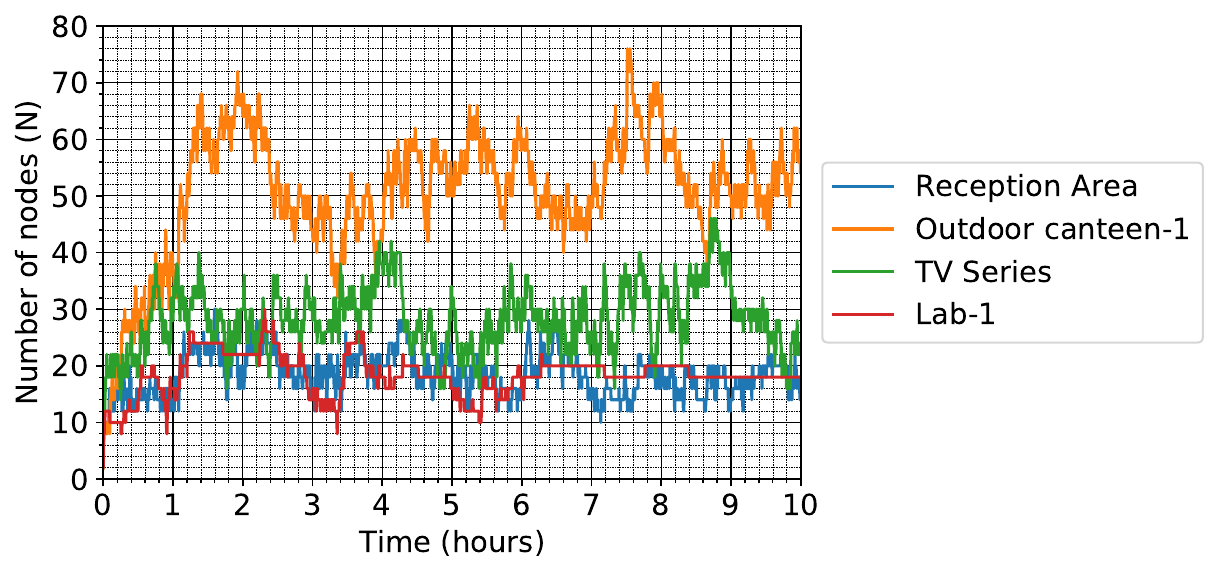}}
\caption{Number of nodes in the tree over a duration of 10 hours in multiple challenging datasets.}
\label{fig:treeHeight}
\end{figure}

Figure~\ref{fig:treeHeight} shows the number of nodes (`N') in the tree over a duration of 10 hours. The `Lab-1' dataset in this plot (red line in the figure) is recorded from 6 PM to 4 AM. Generally, the lab is vacant from midnight to early morning (12 midnight - 5 AM). We can observe that after 6 hours, the number of nodes becomes almost static (approx 20 nodes). It verifies that with no change in the environment, the tree becomes stable. Another important observation is that the number of nodes required in an outdoor environment (`Outdoor Canteen-1' dataset with orange line) is more in comparison to an indoor environment (`Lab-1' dataset). This is because a change in an outdoor scenario is much rapid than in an indoor scenario. The plot for the random audio dataset shows a comparatively increasing graph over time, but it shows convergence later on. We can infer from the plot that even if we use any random audio data, the tree will converge after some time when it has seen enough variations.
\begin{table}[]
\caption{\textcolor{black}{Performance comparison between fixed length Huffman tree and proposed approach on the `Outdoor Canteen-2' dataset. We set $\theta_{cos}$=0.93 and $\alpha$=1e-04 for this experiment.}}
  \label{tab:mainCanteeTable}
\begin{tabular}{ccc|cccc}
\toprule
\multicolumn{3}{c|}{Fixed Length Tree} & \multicolumn{4}{c}{Variable Length Tree}         \\ \hline
N        & Replaced      & AUC(\%)      & AUC(\%) & \multicolumn{1}{c}{avg. N} & Merged & $\theta_{merge}$ \\ \hline
9    & 2046  & 76.57      & 80.94  &119   & 340   & 0.96    \\
11   & 1805 & 75.44          &\textbf{86.00 } & 130  & 235    & 0.97    \\
13  & 1549 & 76.14           & 82.16  &141  & 132   & 0.98    \\
15   & 1329 & \textbf{79.78} & 82.42   & 157  & 50    & 0.99    \\

17   & 1199 & 79.53&&&&\\
19   & 1141&77.53 &&&&\\
\bottomrule
\end{tabular}
\end{table}
\begin{table}[]
\caption{\textcolor{black}{Performance comparison between fixed length Huffman tree and proposed approach on `12 Angry Men' dataset. We set $\theta_{cos}$=0.88 and $\alpha$=1e-04 for this experiment.}}
  \label{tab:12angryMenTable}
\begin{tabular}{ccc|cccc}
\toprule
\multicolumn{3}{c|}{Fixed Length Tree} & \multicolumn{4}{c}{Variable Length Tree}         \\ \hline
N        & Replaced      & AUC(\%)      & AUC(\%) & \multicolumn{1}{c}{avg. N} & Merged & $\theta_{merge}$ \\ \hline
9    & 266   & 61.23 &   80.28  & 14   &64    & 0.96    \\
11   & 151  & \textbf{75.48} &81.34& 16    &44    & 0.97    \\
13  &101 & 69.38  &     \textbf{83.61}  &17  & 22  & 0.98   \\
15   & 57 & 66.61 &      75.44 & 21   &5   & 0.99   \\
17   & 39 & 62.47&&&&\\
19   & 32&74.90 &&&&\\
\bottomrule
\end{tabular}
\end{table}
\begin{table}[]
\caption{\textcolor{black}{Performance comparison between fixed length Huffman tree and proposed approach on `An Outdoor Area' dataset. We set $\theta_{cos}$=0.94 and $\alpha$=1e-03 for this experiment.}}
  \label{tab:kerlaTable}
\begin{tabular}{ccc|cccc}
\toprule
\multicolumn{3}{c|}{Fixed Length Tree} & \multicolumn{4}{c}{Variable Length Tree}   \\ \hline
N  & Replaced   & AUC(\%) & AUC(\%) & \multicolumn{1}{c}{avg. N} &Merged & $\theta_{merge}$ \\ \hline
9    & 191  &58.48  &     57.04  & 26   &134   & 0.96    \\
11   & 162  & 52.49  &     58.40   & 29    &108    & 0.97    \\
13  & 127 & \textbf{58.62} &       \textbf{61.12}  &32 & 76   & 0.98    \\
15   &98 & 48.14 &        60.94  &31   & 50    & 0.99    \\
17   & 80& 42.90&&&&\\
19   & 72&43.48 &&&&\\
\bottomrule
\end{tabular}
\end{table}
\subsection{Fixed length vs. Variable length Tree}
We compare the proposed approach, which has variable tree size, with the fixed-length Huffman tree-based approach on the test volume (1/3 volume) of the three datasets, namely `Outdoor Canteen-2', `12 Angry Men', and `An Outdoor Area'. \textcolor{black}{Similar to a baseline AGMM, when the number of nodes in a fixed-length Huffman exceeds the predefined limit, the node with the smallest weight is replaced by a node containing the new data.}
The comparison results are given in Tables~\ref{tab:mainCanteeTable}, ~\ref{tab:12angryMenTable}, and ~\ref{tab:kerlaTable}, respectively, on the three datasets. 
We keep suitable values of $\theta_{cos}$ and $\alpha$ \textcolor{black}{selected in the hyper-parameter tuning step} for each of the three datasets \textcolor{black}{recorded in different environments}. The finalized values ($\theta_{cos}$, $\alpha$) are as follows: `Outdoor Canteen-2' (0.93, 1e-04), `12 Angry Men' (0.88, 1e-04), `An Outdoor Area' (0.94, 1e-03).
We vary `N' between 9-19 in the fixed Huffman tree method \textcolor{black}{as it is generally kept low in fixed node approaches}. 
We report the number of nodes \textit{replaced} and the AUC measure for this case. For the proposed approach, we mention the \textcolor{black}{average} number of nodes in the tree, the number of total \textit{merged} nodes over total \textcolor{black}{test} dataset duration, and the AUC measure for different values of $\theta_{merge}$. \textcolor{black}{Note that the tree is built on audio frames and hence the reported \textit{replaced} or \textit{merged} count denotes the count of frame instances, not events. There are a total of $8*1798$, $8*880$, and $8*1798$ audio frames for the duration of 1 hour, 30 minutes, and 1 hour of the test volume in `Outdoor Canteen-2', `12 Angry Men', and `An Outdoor Area', respectively.}
From the table, we observe that number of \textit{replaced} nodes decreases as we increase `N' in the fixed Huffman method. It should be so because with a larger `N', the chances of getting a hit increase. The optimal performance is obtained with \textcolor{black}{N=15, 11, and 13 for `Outdoor Canteen-2', `12 Angry Men', and `An Outdoor Area' datasets, respectively}. We can make a similar observation for the proposed approach method with $\theta_{merge}$ and the number of merged nodes. As we increase the merging threshold, nodes are less likely to be merged.
For a higher value of $\theta_{merge}$, two nodes which are indeed similar may be left unmerged.
Whereas a lower $\theta_{merge}$ may end up merging two distinct nodes. 
In the case of the `Outdoor Canteen-2' dataset, we get the best AUC of \textcolor{black}{86.00}\% for the proposed case, whereas, for the fixed Huffman tree, we get the best AUC of \textcolor{black}{79.78}\%. We observe improvement for the other two datasets also in Tables~\ref{tab:12angryMenTable} and ~\ref{tab:kerlaTable}.
\begin{figure}
\centerline{\includegraphics[width=1\linewidth]{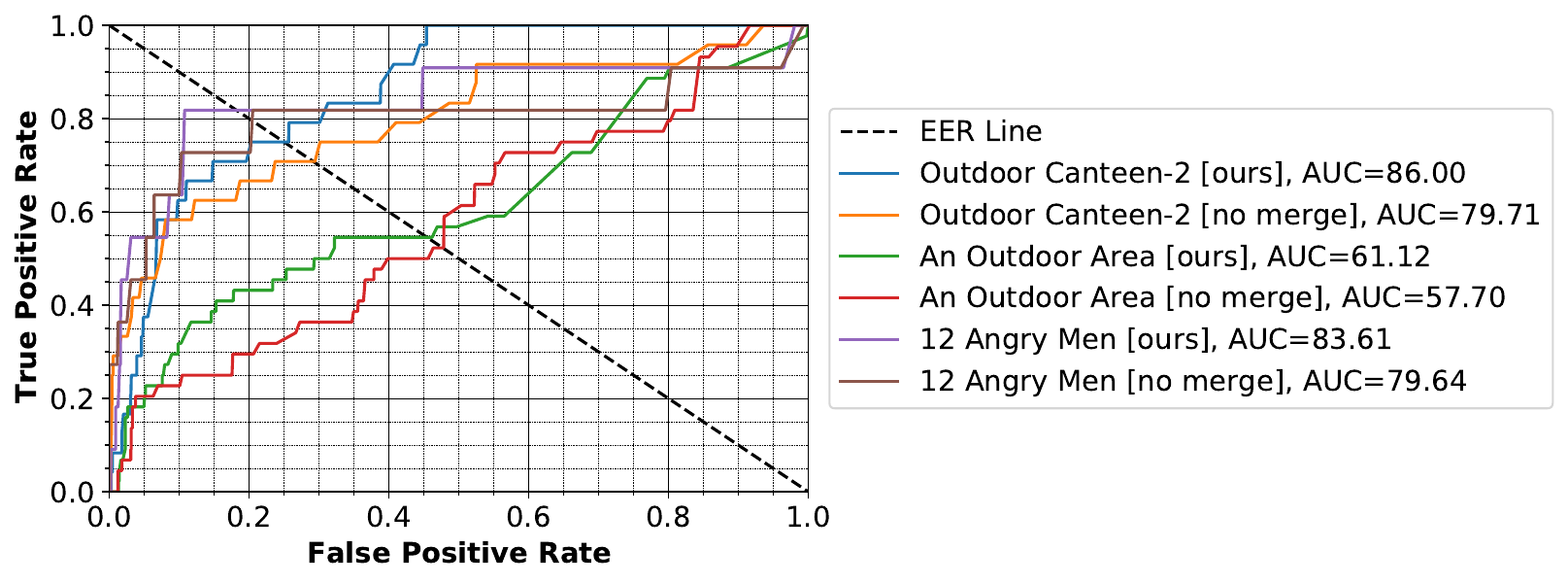}}
\caption{\textcolor{black}{Comparing the performance of the proposed approach with and without the `Merge Nodes' module on `Outdoor Canteen-2', `An Outdoor Area', and `12 angry men' datasets.}}
\label{fig:mergeVSnoMerge}
\end{figure}
\subsection{\textcolor{black}{Merge vs. no Merge in the Adaptive Huffman Tree} }
Here, we discuss the advantage of the `Merge Nodes' module in our approach. The experiments are conducted on the test volume of the three datasets: `An Outdoor Area', `12 Angry Men', and `Outdoor Canteen-2' in Figure~\ref{fig:mergeVSnoMerge}.
We choose the same selected values of $\alpha$ and $\theta_{cos}$ (through hyper-parameter tuning) for both the cases, i.e., merge and no merge. The drastic drop in AUC performance is observed  when we remove the merging module. It signifies the importance of merging redundant nodes to reduces the false alarms.
\begin{table}
   \caption{\textcolor{black}{Comparing hyper-parameters of AGMM and the proposed adaptive Huffman coding based approach.}}
  \label{tab:parmCompare}
  \begin{tabular}{cccccc}
    \toprule
    Description & AGMM & \begin{tabular}[c]{@{}c@{}}Range\\ (AGMM)\end{tabular} & Ours & \begin{tabular}[c]{@{}c@{}}Range\\ (ours)\end{tabular} & Comments \\
    \midrule
    No. of clusters & K &4-6& none &--& adaptive in ours\\
    Merge threshold & none &--& $\theta_{merge}$& 0.96-0.99&extra feature in ours\\
    Weight update& $\alpha_{w}$ &0.001-0.1&$\alpha$& 0.0001-0.1 &required in both\\
    Cluster mean update& $\alpha_{g}$& $\alpha_{w}* \eta$ &$\gamma_{min}$, $\gamma_{max}$&0, 0.5&required in both\\
    Matching threshold & $\theta_{mahalonobis}$ & (2-3)$\sigma$   & $\theta_{cos}$ & 0.88-0.95 & required in both\\
  \bottomrule
\end{tabular}
\end{table}
\subsection{\textcolor{black}{Comparison with the Previous Work }}
As discussed earlier, most of the existing works assume a stationary anomaly distribution, hence, ignore concept drift. We cannot compare the proposed work with those works (supervised anomaly classification) as in our case, the anomaly score for a particular event changes over time due to concept drift. To the best of our knowledge, unsupervised AGMM is the main approach used to address concept drift in audio anomaly detection ~\cite{cristani2004line,cristani2007audio,moncrieff2007online}.
\textcolor{black}{The dynamic scene learning in AGMM and our approach is similar. The mixture in the AGMM evolves over time, i.e., new Gaussians are created to accommodate new samples that are never encountered before. When the mixture reaches the maximum number of possible Gaussians, the weakest Gaussians tend to be replaced by these newly created Gaussians. Along with the mean vector and weight value, the covariance matrices are also maintained for each Gaussian. At the arrival of a new sample, it is matched to the existing Gaussians using Mahalanobis distance. If a match is found, parameters of the matched Gaussian are updated dynamically. Otherwise, a new Gaussian is initialized. After this, weights are normalized too.}

\textcolor{black}{In AGMM, we get hard class labels as abnormal or normal for a fixed value of the foreground-background threshold and hence only one point in ROC space. 
However, to generate a full ROC curve, a discrete classifier (hard-type) can be mapped to a scoring classifier (soft-type) by ``looking inside'' the model~\cite{fawcett2006introduction}. For AGMM, we utilize each Gaussian's weight to convert the binary classifier into a scoring classifier. We sort the Gaussians in decreasing order of weights and use their normalized index (starting from 0) as the anomaly score of each Gaussian. In this way, the Gaussian with maximum weight gets a score of `0', Gaussian with minimum weight gets a score of `1'. 
An audio frame is assigned an anomaly score same as the score of matching or newly added Gaussian. Further, the anomaly score for an event is computed as the average score of all the audio frames in the event. Once we convert the hard label into a soft score, ROC points can be generated by the same algorithm as used for Huffman tree-based approaches.}

\textcolor{black}{Table~\ref{tab:parmCompare} describes the required hyper-parameters along with their range for both of the approaches. We see that the `No. of clusters' parameter is not required for our method. Moreover, the node merging avoids multiple instances of the same event. }
In literature~\cite{zang2006parameter,cristani2004line,cristani2007audio,moncrieff2007online}, the hyper-parameters of the multivariate AGMM are varied as follows: K=3-6 (number of Gaussian), $w_{o}$=0.01-0.1 (initial weight), $\alpha_{w}$=0.001-0.1 (weight update), $\alpha_{g}$=0.001-0.1 (Gaussian update), T=0.92-0.99 (\% data accounted for background events), and $\theta_{mahalonobis}$=4.5 (Z-score is replaced by Mahalanobis distance for multivariate AGMM and for a feature-length of 15 sigma rule says ~4.5$\sigma$ contains 90\% population~\cite{gallego2013mahalanobis} ).

\begin{figure}
\centerline{\includegraphics[width=1\linewidth]{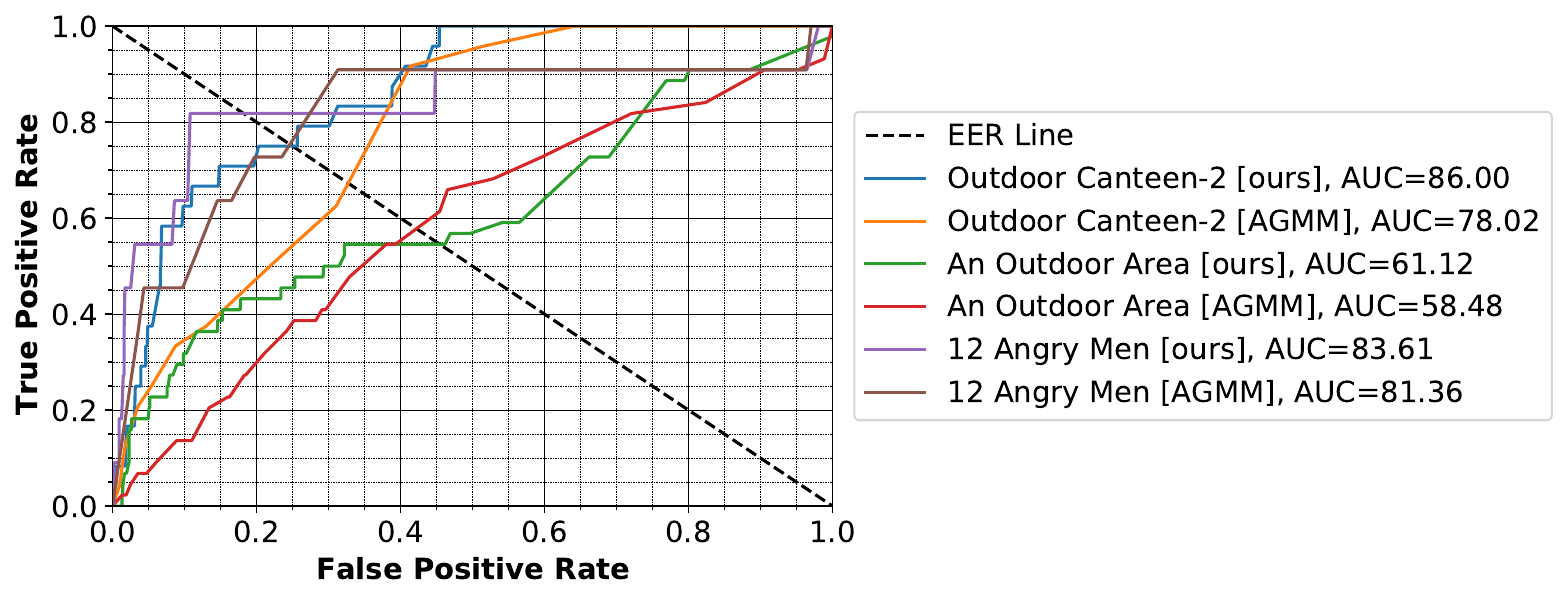}}
\caption{\textcolor{black}{ROC plots for the proposed approach and AGMM on selected hyper-parameters for `Outdoor Canteen-2', `An Outdoor Area', and `12 angry men' datasets.}}
\label{fig:oursVSGMM}
\end{figure}

\textcolor{black}{We compare the AUC performance of our approach with the AGMM approach on different values of `K' (No. of clusters) for each of the three test datasets in Table~\ref{tab:gmmTable}. We can see the proposed approach outperforms the existing AGMM approach on all three datasets. 
The table also reports the number of \textit{replaced} (in the case of AGMM) or \textit{merged} nodes.
We can see that, similar to the fixed Huffman approach, the number of \textit{replaced} nodes is decreasing as we increase `K' in AGMM, i.e., with a larger value of `K', the chances of a miss are less. 
However, the learning and forgetting strategy in the AGMM is causing a large amount of replacements as observed from Table~\ref{tab:gmmTable}. Such a frequent replacement of nodes causes poor memory of the model and hence degradation in the performance.
Another limitation of this approach is the node drifting. The two nodes may drift and come very close to each other, Which means they represent the same class of events. However, their weights may be very different leading to misclassification.
For better visualization, we compare the ROC curve of our approach and the AGMM approach (on `K' with the best performance) in Figure~\ref{fig:oursVSGMM}.}
\begin{table}[]
\caption{\textcolor{black}{Performance comparison between proposed approach and AGMM on the `An Outdoor Area', `12 Angry Men', and `Outdoor Canteen-2' dataset.}}
  \label{tab:gmmTable}
\begin{tabular}{c|cccc}
\toprule
Dataset&Method  & No. of clusters  & AUC(\%)  & Replaced/Merged  \\ 
        \hline
        &AGMM    & 3   & \textbf{78.02}    &2928       \\
         &AGMM    & 4   & 75.87    &2152       \\
Outdoor Canteen-2 &AGMM   & 5 &74.00     &1739    \\
         &AGMM  & 6 & 72.94   &1363     \\
         &Ours   & adaptive & \textbf{86.00}   &235         \\
         \hline
            &AGMM    & 3   & 57.91    &3567        \\
            &AGMM    & 4   & \textbf{58.48 }  &2534        \\
An Outdoor Area &AGMM   & 5 &51.42     &2078      \\
            &AGMM  & 6 &54.41    &1736        \\
            &Ours   & adaptive & \textbf{61.12}   &76         \\
            \hline
         &AGMM    & 3   & 61.36   &3063        \\
         &AGMM    & 4   & 58.30    &2551       \\
12 Angry Men &AGMM   & 5 & \textbf{81.36 }    &2290      \\
         &AGMM  & 6 & 72.60   &1776         \\
        &Ours   & adaptive & \textbf{83.61}   &22         \\
        \bottomrule
\end{tabular}
\end{table}
\begin{figure}
\begin{minipage}[b]{1\linewidth}
  \centering
 \includegraphics[width = 1\linewidth]{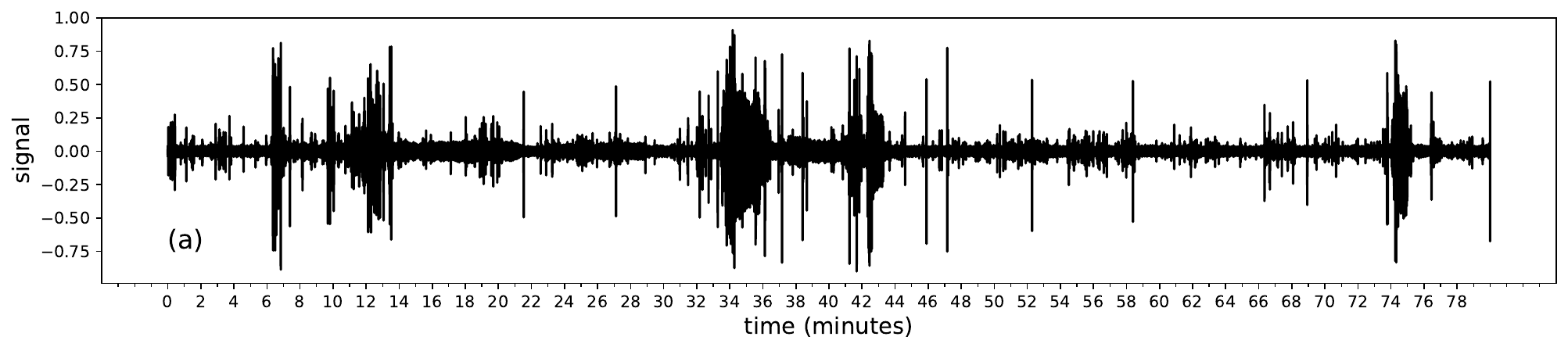}
\end{minipage}

\begin{minipage}[b]{1\linewidth}
  \centering
 \includegraphics[width =1\linewidth]{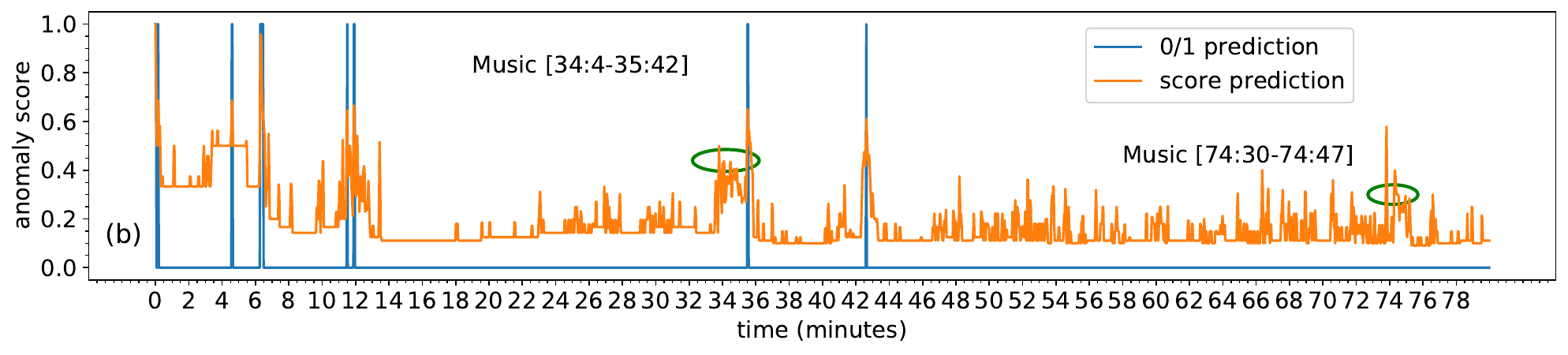}
\end{minipage}
\begin{minipage}[b]{1\linewidth}
  \centering
 \includegraphics[width = 1\linewidth]{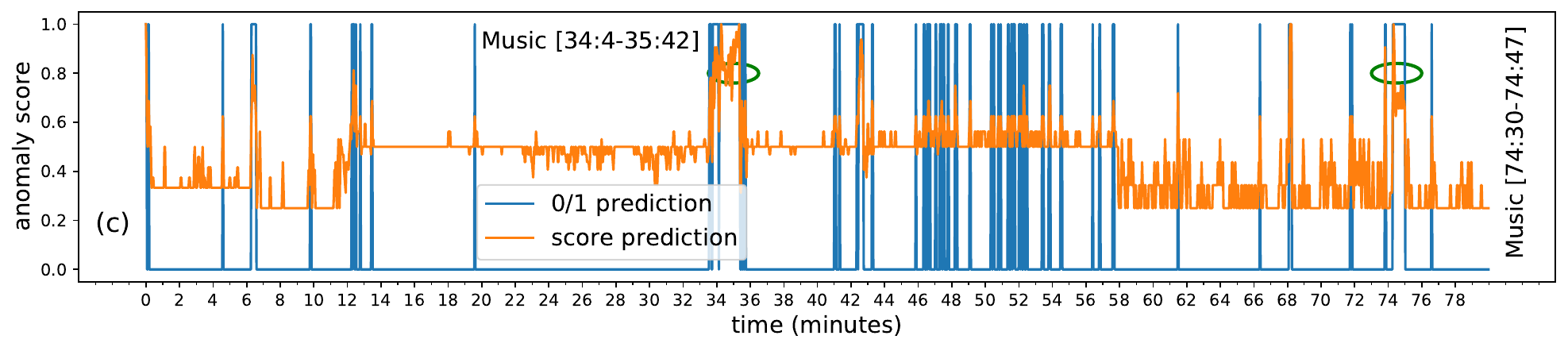}
\end{minipage}
\begin{minipage}[b]{1\linewidth}
  \centering
 \includegraphics[width = 1\linewidth]{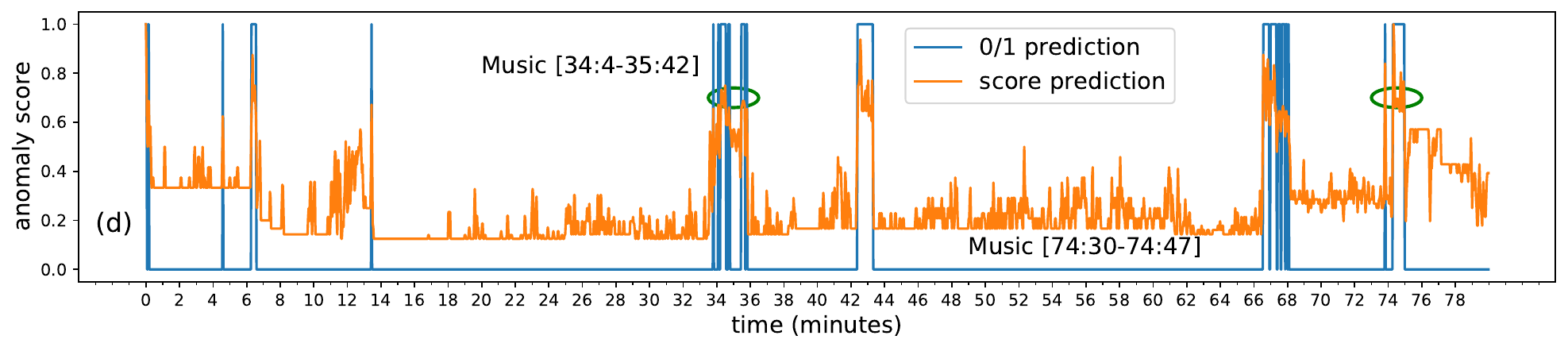}
\end{minipage}
\begin{minipage}[b]{1\linewidth}
  \centering
 \includegraphics[width = 1\linewidth]{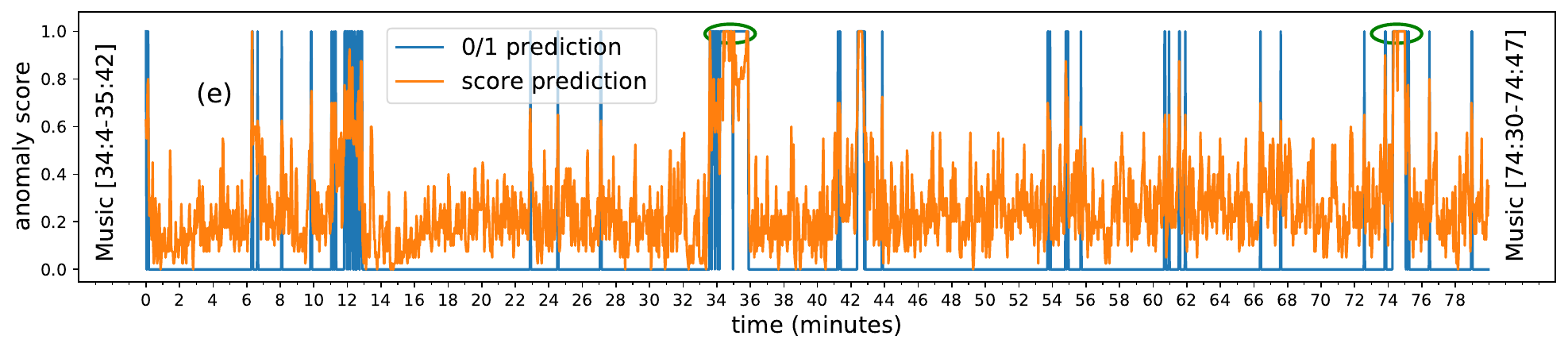}
\end{minipage}
\caption{The plot shows the anomaly score given by the proposed approach, fixed Huffman coding, and AGMM approach for the `Lab-2' dataset. \textcolor{black}{The raw audio is plotted in Fig.(a).} Fig.(b) shows the predicted anomaly scores by the proposed approach. Fig.(c), (d), and (e) show predicted anomaly score by 9-node, 17-node dynamic Huffman Tree, and AGMM, respectively. \textcolor{black}{The parameters for each sub-figures is as follows:
(b) proposed approach ($\theta_{cos}$=0.90, $\theta_{merge}$=0.98, $\alpha$=1e-04, and $w_{o}$=1e-04),
(c) fixed length Huffman tree ($\theta_{cos}$=0.90, N=9, $\alpha$=1e-04, and $w_{o}$=1e-04), 
(d) fixed length Huffman tree ($\theta_{cos}$=0.90, N=17, $\alpha$=1e-04, and $w_{o}$=1e-04), and
 (e) AGMM (K=6, $w_{o}$=0.1, $\alpha_{w}$=0.1, $\alpha_{g}$=0.01, and $\theta_{mahalonobis}$=4.5). }
The dataset contains an event (song played) which is observed multiple times. The proposed approach shows adaptiveness to the new normality. When it sees the event for a significant time, it remembers and produces a low anomaly score. On the other hand, fixed node approaches (c, d, e) that tend to keep short term temporal context forget the long past. Thus when they encounter the same event again, it produces a high score.}
\label{fig:sceneStats}
\end{figure}
\subsection{Scene Adaptiveness}
In this section, we closely analyze what happens when the model encounters the same event again and again. We utilize 1.33 hours long `Lab-2' dataset for the purpose. Figure~\ref{fig:sceneStats}(a) shows the raw audio plot over time. A song was played for the following (start-end) timestamps (minute:second): \{(34:4-35:42), (74:30-74:47)\}. 
We have marked these stamps on the plots. Figure~\ref{fig:sceneStats}(b) depicts the raw anomaly score (orange color in the figure) as well as 0/1 decision (black color in the figure) for the proposed approach. \textcolor{black}{The threshold to convert raw score in 0/1 prediction is kept 0.6 (this is just for the plot, it does not affect any inference in this experiment).}
The best suitable hyper-parameters for this dataset are $\theta_{cos}$=0.90, $\theta_{merge}$=0.98, and $\alpha$=1e-04. The number of total nodes reaches up-to 17 in this case. We plot the same for the fixed-length Huffman coding. Figure~\ref{fig:sceneStats}(c) and \ref{fig:sceneStats}(d) represent the anomaly scores for a fixed length of 9 and 17 nodes, respectively. Figure~\ref{fig:sceneStats}(e) shows the raw score as well as 0/1 predictions from
AGMM with the hyper-parameters as K=6, $w_{o}$=0.1, $\alpha_{w}$=0.1, $\alpha_{g}$=0.01, and $\theta_{mahalonobis}$=4.5. From the above plots, it can be observed that the anomaly score gradually decreases in the interval \{34:4-35:42\} for the proposed case. It should decrease because the node corresponding to the song is getting enough neighbors, and hence the node attains a lower depth in the tree. The same behavior can be observed in the fixed node cases also (Figure~\ref{fig:sceneStats}(c), (d), (e)). Whether we use the fixed length or variable length approaches, the online models are able to adapt to the short term temporal context. But when the K-node models encounter the same event (the same song played in interval 74:30-74:47) after a long duration, it again produces a high anomaly score and classifies it as abnormal. This is because models with fixed `N'  \textcolor{black}{have a greater tendency to forget the distant past than the proposed approach does. In both approaches, weight normalization results in a gradual forgetting of the events. However, with fixed nodes, you may completely forget the past if the node is replaced. In contrast, our model remembers a longer history, although  with a reduced weight depending on the frequency of occurrence. In the case above, our method initially puts the node corresponding to this event in the lower part, but gradually the node moves up in the tree because the same event was observed for a large duration (> 95 seconds). The last occurrence of the  event gets a low anomaly score due to the high weight of the node. During the interval when the event is not observed, the weight of the node reduces, but not by much amount. Therefore, the second time the event occurs, the model assigns a relatively low anomaly score to the event, as shown in Figure~\ref{fig:sceneStats}(b).}
\section{Limitations}\label{sec:limitations}
\textcolor{black}{The proposed model is unsupervised; it trains itself automatically with time. In order to train the model, the data samples (frames in our case) need to arrive in the temporal order. Therefore, the model cannot be trained on traditional datasets where temporal order is missing (such as DCASE \cite{DCASE2017challenge}). Another limitation of the work is that the hyper-parameters need to be tuned according to the application scenario. However, it is only a one-time task, and the same set of parameters should work in all similar scenarios. Finally, like all previous multimedia based anomaly detection works, we fail to accommodate the periodicity of the events; we are only able to track the trend in the data.  }
\section{CONCLUSION}\label{sec:conc}
The proposed framework is able to effectively apply adaptive Huffman coding for anomaly detection in audio data having concept drift. The dynamic tree construction and reorganization are able to accommodate concept drift in audio. New data adapts the codes representing the leaf nodes. Instead of replacing the old nodes, we propose to merge the nodes when they are close enough. The proposed method with node merging framework is able to grasp the long past, while the fixed node approaches with replacement policy forget the long term history. The experiments also show that the number of maximum nodes remains tractable; it is found that the number of nodes converges between 70 and 75 for most continuous audio datasets, even when the audio length is very long. The proposed method works effectively on three datasets, referring to challenging outdoor scenarios. We are able to outperform an AGMM framework that requires prior knowledge about data distribution. In the future, we would like to extend the work to include video information to build the Huffman tree and detect the anomaly. 

\bibliographystyle{ACM-Reference-Format}
\bibliography{main}


\begin{thebibliography}{47}


\ifx \showCODEN    \undefined \def \showCODEN     #1{\unskip}     \fi
\ifx \showDOI      \undefined \def \showDOI       #1{#1}\fi
\ifx \showISBNx    \undefined \def \showISBNx     #1{\unskip}     \fi
\ifx \showISBNxiii \undefined \def \showISBNxiii  #1{\unskip}     \fi
\ifx \showISSN     \undefined \def \showISSN      #1{\unskip}     \fi
\ifx \showLCCN     \undefined \def \showLCCN      #1{\unskip}     \fi
\ifx \shownote     \undefined \def \shownote      #1{#1}          \fi
\ifx \showarticletitle \undefined \def \showarticletitle #1{#1}   \fi
\ifx \showURL      \undefined \def \showURL       {\relax}        \fi
\providecommand\bibfield[2]{#2}
\providecommand\bibinfo[2]{#2}
\providecommand\natexlab[1]{#1}
\providecommand\showeprint[2][]{arXiv:#2}

\bibitem[\protect\citeauthoryear{Ahmad, Lavin, Purdy, and Agha}{Ahmad
  et~al\mbox{.}}{2017}]%
        {ahmad2017unsupervised}
\bibfield{author}{\bibinfo{person}{Subutai Ahmad}, \bibinfo{person}{Alexander
  Lavin}, \bibinfo{person}{Scott Purdy}, {and} \bibinfo{person}{Zuha Agha}.}
  \bibinfo{year}{2017}\natexlab{}.
\newblock \showarticletitle{Unsupervised real-time anomaly detection for
  streaming data}.
\newblock \bibinfo{journal}{\emph{Neurocomputing}}  \bibinfo{volume}{262}
  (\bibinfo{year}{2017}), \bibinfo{pages}{134--147}.
\newblock


\bibitem[\protect\citeauthoryear{Ahmad and Purdy}{Ahmad and Purdy}{2016}]%
        {ahmad2016real}
\bibfield{author}{\bibinfo{person}{Subutai Ahmad} {and} \bibinfo{person}{Scott
  Purdy}.} \bibinfo{year}{2016}\natexlab{}.
\newblock \showarticletitle{Real-time anomaly detection for streaming
  analytics}.
\newblock \bibinfo{journal}{\emph{arXiv preprint arXiv:1607.02480}}
  (\bibinfo{year}{2016}).
\newblock


\bibitem[\protect\citeauthoryear{Basseville, Nikiforov,
  et~al\mbox{.}}{Basseville et~al\mbox{.}}{1993}]%
        {basseville1993detection}
\bibfield{author}{\bibinfo{person}{Michele Basseville}, \bibinfo{person}{Igor~V
  Nikiforov}, {et~al\mbox{.}}} \bibinfo{year}{1993}\natexlab{}.
\newblock \bibinfo{booktitle}{\emph{Detection of abrupt changes: theory and
  application}}. Vol.~\bibinfo{volume}{104}.
\newblock \bibinfo{publisher}{Prentice hall Englewood Cliffs}.
\newblock


\bibitem[\protect\citeauthoryear{Bianco, Garcia~Ben, Martinez, and
  Yohai}{Bianco et~al\mbox{.}}{2001}]%
        {bianco2001outlier}
\bibfield{author}{\bibinfo{person}{Ana~Maria Bianco}, \bibinfo{person}{M
  Garcia~Ben}, \bibinfo{person}{EJ Martinez}, {and}
  \bibinfo{person}{V{\i}ctor~J Yohai}.} \bibinfo{year}{2001}\natexlab{}.
\newblock \showarticletitle{Outlier detection in regression models with arima
  errors using robust estimates}.
\newblock \bibinfo{journal}{\emph{Journal of Forecasting}}
  \bibinfo{volume}{20}, \bibinfo{number}{8} (\bibinfo{year}{2001}),
  \bibinfo{pages}{565--579}.
\newblock


\bibitem[\protect\citeauthoryear{B{\"o}hm, Haegler, M{\"u}ller, and
  Plant}{B{\"o}hm et~al\mbox{.}}{2009}]%
        {bohm2009coco}
\bibfield{author}{\bibinfo{person}{Christian B{\"o}hm}, \bibinfo{person}{Katrin
  Haegler}, \bibinfo{person}{Nikola~S M{\"u}ller}, {and}
  \bibinfo{person}{Claudia Plant}.} \bibinfo{year}{2009}\natexlab{}.
\newblock \showarticletitle{CoCo: coding cost for parameter-free outlier
  detection}. In \bibinfo{booktitle}{\emph{Proceedings of the 15th SIGKDD
  international conference on Knowledge discovery and data mining}}.
  \bibinfo{publisher}{ACM}, \bibinfo{address}{Paris, France},
  \bibinfo{pages}{149--158}.
\newblock


\bibitem[\protect\citeauthoryear{Bollegala}{Bollegala}{2017}]%
        {bollegala2017dynamic}
\bibfield{author}{\bibinfo{person}{Danushka Bollegala}.}
  \bibinfo{year}{2017}\natexlab{}.
\newblock \showarticletitle{Dynamic feature scaling for online learning of
  binary classifiers}.
\newblock \bibinfo{journal}{\emph{Knowledge-Based Systems}}
  \bibinfo{volume}{129} (\bibinfo{year}{2017}), \bibinfo{pages}{97--105}.
\newblock


\bibitem[\protect\citeauthoryear{Bradley}{Bradley}{1997}]%
        {bradley1997use}
\bibfield{author}{\bibinfo{person}{Andrew~P Bradley}.}
  \bibinfo{year}{1997}\natexlab{}.
\newblock \showarticletitle{The use of the area under the ROC curve in the
  evaluation of machine learning algorithms}.
\newblock \bibinfo{journal}{\emph{Pattern recognition}} \bibinfo{volume}{30},
  \bibinfo{number}{7} (\bibinfo{year}{1997}), \bibinfo{pages}{1145--1159}.
\newblock


\bibitem[\protect\citeauthoryear{Callegari, Giordano, and Pagano}{Callegari
  et~al\mbox{.}}{2009}]%
        {callegari2009use}
\bibfield{author}{\bibinfo{person}{Christian Callegari},
  \bibinfo{person}{Stefano Giordano}, {and} \bibinfo{person}{Michele Pagano}.}
  \bibinfo{year}{2009}\natexlab{}.
\newblock \showarticletitle{On the use of compression algorithms for network
  anomaly detection}. In \bibinfo{booktitle}{\emph{International Conference on
  Communications}}. \bibinfo{publisher}{IEEE}, \bibinfo{address}{Dresden,
  Germany}, \bibinfo{pages}{1--5}.
\newblock


\bibitem[\protect\citeauthoryear{Carletti, Foggia, Percannella, Saggese,
  Strisciuglio, and Vento}{Carletti et~al\mbox{.}}{2013}]%
        {carletti2013audio}
\bibfield{author}{\bibinfo{person}{Vincenzo Carletti},
  \bibinfo{person}{Pasquale Foggia}, \bibinfo{person}{Gennaro Percannella},
  \bibinfo{person}{Alessia Saggese}, \bibinfo{person}{Nicola Strisciuglio},
  {and} \bibinfo{person}{Mario Vento}.} \bibinfo{year}{2013}\natexlab{}.
\newblock \showarticletitle{Audio surveillance using a bag of aural words
  classifier}. In \bibinfo{booktitle}{\emph{10th International Conference on
  Advanced Video and Signal Based Surveillance}}. \bibinfo{publisher}{IEEE},
  \bibinfo{address}{Krakow, Poland}, \bibinfo{pages}{81--86}.
\newblock


\bibitem[\protect\citeauthoryear{Cohen}{Cohen}{1960}]%
        {cohen1960coefficient}
\bibfield{author}{\bibinfo{person}{Jacob Cohen}.}
  \bibinfo{year}{1960}\natexlab{}.
\newblock \showarticletitle{A coefficient of agreement for nominal scales}.
\newblock \bibinfo{journal}{\emph{Educational and psychological measurement}}
  \bibinfo{volume}{20}, \bibinfo{number}{1} (\bibinfo{year}{1960}),
  \bibinfo{pages}{37--46}.
\newblock


\bibitem[\protect\citeauthoryear{Conte, Foggia, Percannella, Saggese, and
  Vento}{Conte et~al\mbox{.}}{2012}]%
        {conte2012ensemble}
\bibfield{author}{\bibinfo{person}{Donatello Conte}, \bibinfo{person}{Pasquale
  Foggia}, \bibinfo{person}{Gennaro Percannella}, \bibinfo{person}{Alessia
  Saggese}, {and} \bibinfo{person}{Mario Vento}.}
  \bibinfo{year}{2012}\natexlab{}.
\newblock \showarticletitle{An ensemble of rejecting classifiers for anomaly
  detection of audio events}. In \bibinfo{booktitle}{\emph{Ninth International
  Conference on Advanced Video and Signal-Based Surveillance}}.
  \bibinfo{publisher}{IEEE}, \bibinfo{address}{Beijing, China},
  \bibinfo{pages}{76--81}.
\newblock


\bibitem[\protect\citeauthoryear{Costa, Silva, Antunes, and Ribeiro}{Costa
  et~al\mbox{.}}{2014}]%
        {costa2014concept}
\bibfield{author}{\bibinfo{person}{Joana Costa}, \bibinfo{person}{Catarina
  Silva}, \bibinfo{person}{M{\'a}rio Antunes}, {and}
  \bibinfo{person}{Bernardete Ribeiro}.} \bibinfo{year}{2014}\natexlab{}.
\newblock \showarticletitle{Concept drift awareness in twitter streams}. In
  \bibinfo{booktitle}{\emph{13th International Conference on Machine Learning
  and Applications}}. \bibinfo{publisher}{IEEE}, \bibinfo{address}{Detroit, MI,
  USA}, \bibinfo{pages}{294--299}.
\newblock


\bibitem[\protect\citeauthoryear{Cristani, Bicego, and Murino}{Cristani
  et~al\mbox{.}}{2004}]%
        {cristani2004line}
\bibfield{author}{\bibinfo{person}{Marco Cristani}, \bibinfo{person}{Manuele
  Bicego}, {and} \bibinfo{person}{Vittorio Murino}.}
  \bibinfo{year}{2004}\natexlab{}.
\newblock \showarticletitle{On-line adaptive background modelling for audio
  surveillance}. In \bibinfo{booktitle}{\emph{Proceedings of the 17th
  International Conference on Pattern Recognition}}. \bibinfo{publisher}{IEEE},
  \bibinfo{address}{Cambridge, UK}, \bibinfo{pages}{399--402}.
\newblock


\bibitem[\protect\citeauthoryear{Cristani, Bicego, and Murino}{Cristani
  et~al\mbox{.}}{2007}]%
        {cristani2007audio}
\bibfield{author}{\bibinfo{person}{Marco Cristani}, \bibinfo{person}{Manuele
  Bicego}, {and} \bibinfo{person}{Vittorio Murino}.}
  \bibinfo{year}{2007}\natexlab{}.
\newblock \showarticletitle{Audio-visual event recognition in surveillance
  video sequences}.
\newblock \bibinfo{journal}{\emph{Transactions on Multimedia}}
  \bibinfo{volume}{9}, \bibinfo{number}{2} (\bibinfo{year}{2007}),
  \bibinfo{pages}{257--267}.
\newblock


\bibitem[\protect\citeauthoryear{Crocco, Cristani, Trucco, and Murino}{Crocco
  et~al\mbox{.}}{2016}]%
        {crocco2016audio}
\bibfield{author}{\bibinfo{person}{Marco Crocco}, \bibinfo{person}{Marco
  Cristani}, \bibinfo{person}{Andrea Trucco}, {and} \bibinfo{person}{Vittorio
  Murino}.} \bibinfo{year}{2016}\natexlab{}.
\newblock \showarticletitle{Audio surveillance: A systematic review}.
\newblock \bibinfo{journal}{\emph{Comput. Surveys}} \bibinfo{volume}{48},
  \bibinfo{number}{4} (\bibinfo{year}{2016}), \bibinfo{pages}{1--46}.
\newblock


\bibitem[\protect\citeauthoryear{Fawcett}{Fawcett}{2006}]%
        {fawcett2006introduction}
\bibfield{author}{\bibinfo{person}{Tom Fawcett}.}
  \bibinfo{year}{2006}\natexlab{}.
\newblock \showarticletitle{An introduction to ROC analysis}.
\newblock \bibinfo{journal}{\emph{Pattern recognition letters}}
  \bibinfo{volume}{27}, \bibinfo{number}{8} (\bibinfo{year}{2006}),
  \bibinfo{pages}{861--874}.
\newblock


\bibitem[\protect\citeauthoryear{Fenza, Gallo, and Loia}{Fenza
  et~al\mbox{.}}{2019}]%
        {fenza2019drift}
\bibfield{author}{\bibinfo{person}{Giuseppe Fenza},
  \bibinfo{person}{Mariacristina Gallo}, {and} \bibinfo{person}{Vincenzo
  Loia}.} \bibinfo{year}{2019}\natexlab{}.
\newblock \showarticletitle{Drift-aware methodology for anomaly detection in
  smart grid}.
\newblock \bibinfo{journal}{\emph{IEEE Access}}  \bibinfo{volume}{7}
  (\bibinfo{year}{2019}), \bibinfo{pages}{9645--9657}.
\newblock


\bibitem[\protect\citeauthoryear{Foggia, Petkov, Saggese, Strisciuglio, and
  Vento}{Foggia et~al\mbox{.}}{2015a}]%
        {foggia2015audio}
\bibfield{author}{\bibinfo{person}{Pasquale Foggia}, \bibinfo{person}{Nicolai
  Petkov}, \bibinfo{person}{Alessia Saggese}, \bibinfo{person}{Nicola
  Strisciuglio}, {and} \bibinfo{person}{Mario Vento}.}
  \bibinfo{year}{2015}\natexlab{a}.
\newblock \showarticletitle{Audio surveillance of roads: A system for detecting
  anomalous sounds}.
\newblock \bibinfo{journal}{\emph{Transactions on intelligent transportation
  systems}} \bibinfo{volume}{17}, \bibinfo{number}{1} (\bibinfo{year}{2015}),
  \bibinfo{pages}{279--288}.
\newblock


\bibitem[\protect\citeauthoryear{Foggia, Petkov, Saggese, Strisciuglio, and
  Vento}{Foggia et~al\mbox{.}}{2015b}]%
        {foggia2015reliable}
\bibfield{author}{\bibinfo{person}{Pasquale Foggia}, \bibinfo{person}{Nicolai
  Petkov}, \bibinfo{person}{Alessia Saggese}, \bibinfo{person}{Nicola
  Strisciuglio}, {and} \bibinfo{person}{Mario Vento}.}
  \bibinfo{year}{2015}\natexlab{b}.
\newblock \showarticletitle{Reliable detection of audio events in highly noisy
  environments}.
\newblock \bibinfo{journal}{\emph{Pattern Recognition Letters}}
  \bibinfo{volume}{65} (\bibinfo{year}{2015}), \bibinfo{pages}{22--28}.
\newblock


\bibitem[\protect\citeauthoryear{Gallego, Cuevas, Mohedano, and Garcia}{Gallego
  et~al\mbox{.}}{2013}]%
        {gallego2013mahalanobis}
\bibfield{author}{\bibinfo{person}{Guillermo Gallego}, \bibinfo{person}{Carlos
  Cuevas}, \bibinfo{person}{Raul Mohedano}, {and} \bibinfo{person}{Narciso
  Garcia}.} \bibinfo{year}{2013}\natexlab{}.
\newblock \showarticletitle{On the Mahalanobis distance classification
  criterion for multidimensional normal distributions}.
\newblock \bibinfo{journal}{\emph{Transactions on Signal Processing}}
  \bibinfo{volume}{61}, \bibinfo{number}{17} (\bibinfo{year}{2013}),
  \bibinfo{pages}{4387--4396}.
\newblock


\bibitem[\protect\citeauthoryear{Gama, {\v{Z}}liobait{\.e}, Bifet, Pechenizkiy,
  and Bouchachia}{Gama et~al\mbox{.}}{2014}]%
        {gama2014survey}
\bibfield{author}{\bibinfo{person}{Jo{\~a}o Gama}, \bibinfo{person}{Indr{\.e}
  {\v{Z}}liobait{\.e}}, \bibinfo{person}{Albert Bifet}, \bibinfo{person}{Mykola
  Pechenizkiy}, {and} \bibinfo{person}{Abdelhamid Bouchachia}.}
  \bibinfo{year}{2014}\natexlab{}.
\newblock \showarticletitle{A survey on concept drift adaptation}.
\newblock \bibinfo{journal}{\emph{ACM computing surveys}} \bibinfo{volume}{46},
  \bibinfo{number}{4} (\bibinfo{year}{2014}), \bibinfo{pages}{1--37}.
\newblock


\bibitem[\protect\citeauthoryear{Gilligan}{Gilligan}{2009}]%
        {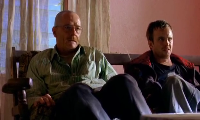}
\bibfield{author}{\bibinfo{person}{Vince Gilligan}.}
  \bibinfo{year}{2009}\natexlab{}.
\newblock \bibinfo{title}{TV Series: Breaking Bad}.
\newblock \bibinfo{howpublished}{\textit{High Bridge Entertainment, Gran Via
  Productions, Sony Pictures Home Entertainment}}.
\newblock


\bibitem[\protect\citeauthoryear{Huffman}{Huffman}{1952}]%
        {huffman1952method}
\bibfield{author}{\bibinfo{person}{David~A Huffman}.}
  \bibinfo{year}{1952}\natexlab{}.
\newblock \showarticletitle{A method for the construction of minimum-redundancy
  codes}.
\newblock \bibinfo{journal}{\emph{Proceedings of the IRE}}
  \bibinfo{volume}{40}, \bibinfo{number}{9} (\bibinfo{year}{1952}),
  \bibinfo{pages}{1098--1101}.
\newblock


\bibitem[\protect\citeauthoryear{Knuth}{Knuth}{1985}]%
        {knuth1985dynamic}
\bibfield{author}{\bibinfo{person}{Donald~E Knuth}.}
  \bibinfo{year}{1985}\natexlab{}.
\newblock \showarticletitle{Dynamic huffman coding}.
\newblock \bibinfo{journal}{\emph{Journal of algorithms}} \bibinfo{volume}{6},
  \bibinfo{number}{2} (\bibinfo{year}{1985}), \bibinfo{pages}{163--180}.
\newblock


\bibitem[\protect\citeauthoryear{Kratz and Nishino}{Kratz and Nishino}{2009}]%
        {kratz2009anomaly}
\bibfield{author}{\bibinfo{person}{Louis Kratz} {and} \bibinfo{person}{Ko
  Nishino}.} \bibinfo{year}{2009}\natexlab{}.
\newblock \showarticletitle{Anomaly detection in extremely crowded scenes using
  spatio-temporal motion pattern models}. In
  \bibinfo{booktitle}{\emph{Conference on Computer Vision and Pattern
  Recognition}}. \bibinfo{publisher}{IEEE}, \bibinfo{address}{Miami, FL, USA},
  \bibinfo{pages}{1446--1453}.
\newblock


\bibitem[\protect\citeauthoryear{Landis and Koch}{Landis and Koch}{1977}]%
        {landis1977measurement}
\bibfield{author}{\bibinfo{person}{J~Richard Landis} {and}
  \bibinfo{person}{Gary~G Koch}.} \bibinfo{year}{1977}\natexlab{}.
\newblock \showarticletitle{The measurement of observer agreement for
  categorical data}.
\newblock \bibinfo{journal}{\emph{biometrics}} \bibinfo{volume}{33},
  \bibinfo{number}{1} (\bibinfo{year}{1977}), \bibinfo{pages}{159--174}.
\newblock


\bibitem[\protect\citeauthoryear{Lim, Kim, and Kim}{Lim et~al\mbox{.}}{2015}]%
        {lim2015robust}
\bibfield{author}{\bibinfo{person}{Hyungjun Lim}, \bibinfo{person}{Myung~Jong
  Kim}, {and} \bibinfo{person}{Hoirin Kim}.} \bibinfo{year}{2015}\natexlab{}.
\newblock \showarticletitle{Robust sound event classification using LBP-HOG
  based bag-of-audio-words feature representation}. In
  \bibinfo{booktitle}{\emph{16th Annual Conference of the International Speech
  Communication Association}}. \bibinfo{publisher}{ISCA},
  \bibinfo{address}{Dresden, Germany}, \bibinfo{pages}{3325--3329}.
\newblock


\bibitem[\protect\citeauthoryear{Lu, Liu, Song, and Zhang}{Lu
  et~al\mbox{.}}{2020}]%
        {lu2020data}
\bibfield{author}{\bibinfo{person}{Jie Lu}, \bibinfo{person}{Anjin Liu},
  \bibinfo{person}{Yiliao Song}, {and} \bibinfo{person}{Guangquan Zhang}.}
  \bibinfo{year}{2020}\natexlab{}.
\newblock \showarticletitle{Data-driven decision support under concept drift in
  streamed big data}.
\newblock \bibinfo{journal}{\emph{Complex \& Intelligent Systems}}
  \bibinfo{volume}{6}, \bibinfo{number}{1} (\bibinfo{year}{2020}),
  \bibinfo{pages}{157--163}.
\newblock


\bibitem[\protect\citeauthoryear{Lumet, Fonda, and Rose}{Lumet
  et~al\mbox{.}}{1957}]%
        {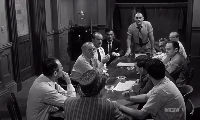}
\bibfield{author}{\bibinfo{person}{Sidney Lumet}, \bibinfo{person}{Henry
  Fonda}, {and} \bibinfo{person}{Reginald Rose}.}
  \bibinfo{year}{1957}\natexlab{}.
\newblock \bibinfo{title}{Movie: Twelve Angry Men}.
\newblock \bibinfo{howpublished}{Orion-Nova Productions}.
\newblock


\bibitem[\protect\citeauthoryear{Ma, Zhang, Pei, Huang, and Dai}{Ma
  et~al\mbox{.}}{2018}]%
        {ma2018robust}
\bibfield{author}{\bibinfo{person}{Minghua Ma}, \bibinfo{person}{Shenglin
  Zhang}, \bibinfo{person}{Dan Pei}, \bibinfo{person}{Xin Huang}, {and}
  \bibinfo{person}{Hongwei Dai}.} \bibinfo{year}{2018}\natexlab{}.
\newblock \showarticletitle{Robust and rapid adaption for concept drift in
  software system anomaly detection}. In \bibinfo{booktitle}{\emph{29th
  International Symposium on Software Reliability Engineering}}.
  \bibinfo{publisher}{IEEE}, \bibinfo{address}{Memphis, TN, USA},
  \bibinfo{pages}{13--24}.
\newblock


\bibitem[\protect\citeauthoryear{Mesaros, Heittola, Diment, Elizalde, Shah,
  Vincent, Raj, and Virtanen}{Mesaros et~al\mbox{.}}{2017}]%
        {DCASE2017challenge}
\bibfield{author}{\bibinfo{person}{A. Mesaros}, \bibinfo{person}{T. Heittola},
  \bibinfo{person}{A. Diment}, \bibinfo{person}{B. Elizalde},
  \bibinfo{person}{A. Shah}, \bibinfo{person}{E. Vincent}, \bibinfo{person}{B.
  Raj}, {and} \bibinfo{person}{T. Virtanen}.} \bibinfo{year}{2017}\natexlab{}.
\newblock \showarticletitle{{DCASE} 2017 Challenge Setup: Tasks, Datasets and
  Baseline System}. In \bibinfo{booktitle}{\emph{Proceedings of the Detection
  and Classification of Acoustic Scenes and Events Workshop}}.
  \bibinfo{publisher}{Inria}, \bibinfo{address}{Nancy, France},
  \bibinfo{pages}{85--92}.
\newblock


\bibitem[\protect\citeauthoryear{Moncrieff, Venkatesh, and West}{Moncrieff
  et~al\mbox{.}}{2007}]%
        {moncrieff2007online}
\bibfield{author}{\bibinfo{person}{Simon Moncrieff}, \bibinfo{person}{Svetha
  Venkatesh}, {and} \bibinfo{person}{Geoff West}.}
  \bibinfo{year}{2007}\natexlab{}.
\newblock \showarticletitle{Online audio background determination for complex
  audio environments}.
\newblock \bibinfo{journal}{\emph{Transactions on Multimedia Computing,
  Communications, and Applications}} \bibinfo{volume}{3}, \bibinfo{number}{2}
  (\bibinfo{year}{2007}), \bibinfo{pages}{8--es}.
\newblock


\bibitem[\protect\citeauthoryear{Roshtkhari and Levine}{Roshtkhari and
  Levine}{2013}]%
        {roshtkhari2013line}
\bibfield{author}{\bibinfo{person}{Mehrsan~Javan Roshtkhari} {and}
  \bibinfo{person}{Martin~D Levine}.} \bibinfo{year}{2013}\natexlab{}.
\newblock \showarticletitle{An on-line, real-time learning method for detecting
  anomalies in videos using spatio-temporal compositions}.
\newblock \bibinfo{journal}{\emph{Computer vision and image understanding}}
  \bibinfo{volume}{117}, \bibinfo{number}{10} (\bibinfo{year}{2013}),
  \bibinfo{pages}{1436--1452}.
\newblock


\bibitem[\protect\citeauthoryear{Rushe and Mac~Namee}{Rushe and
  Mac~Namee}{2019}]%
        {rushe2019anomaly}
\bibfield{author}{\bibinfo{person}{Ellen Rushe} {and} \bibinfo{person}{Brian
  Mac~Namee}.} \bibinfo{year}{2019}\natexlab{}.
\newblock \showarticletitle{Anomaly Detection in Raw Audio Using Deep
  Autoregressive Networks}. In \bibinfo{booktitle}{\emph{International
  Conference on Acoustics, Speech and Signal Processing}}.
  \bibinfo{publisher}{IEEE}, \bibinfo{address}{Brighton, United Kingdom},
  \bibinfo{pages}{3597--3601}.
\newblock


\bibitem[\protect\citeauthoryear{Saurav, Malhotra, TV, Gugulothu, Vig, Agarwal,
  and Shroff}{Saurav et~al\mbox{.}}{2018}]%
        {saurav2018online}
\bibfield{author}{\bibinfo{person}{Sakti Saurav}, \bibinfo{person}{Pankaj
  Malhotra}, \bibinfo{person}{Vishnu TV}, \bibinfo{person}{Narendhar
  Gugulothu}, \bibinfo{person}{Lovekesh Vig}, \bibinfo{person}{Puneet Agarwal},
  {and} \bibinfo{person}{Gautam Shroff}.} \bibinfo{year}{2018}\natexlab{}.
\newblock \showarticletitle{Online anomaly detection with concept drift
  adaptation using recurrent neural networks}. In
  \bibinfo{booktitle}{\emph{Proceedings of the ACM India Joint International
  Conference on Data Science and Management of Data}}.
  \bibinfo{publisher}{ACM}, \bibinfo{address}{Goa, India},
  \bibinfo{pages}{78--87}.
\newblock


\bibitem[\protect\citeauthoryear{Smeaton and McHugh}{Smeaton and
  McHugh}{2005}]%
        {smeaton2005towards}
\bibfield{author}{\bibinfo{person}{Alan~F Smeaton} {and} \bibinfo{person}{Mike
  McHugh}.} \bibinfo{year}{2005}\natexlab{}.
\newblock \showarticletitle{Towards event detection in an audio-based sensor
  network}. In \bibinfo{booktitle}{\emph{Proceedings of the third international
  workshop on Video surveillance \& sensor networks}}.
  \bibinfo{publisher}{ACM}, \bibinfo{address}{Hilton, Singapore},
  \bibinfo{pages}{87--94}.
\newblock


\bibitem[\protect\citeauthoryear{Stauffer and Grimson}{Stauffer and
  Grimson}{1999}]%
        {stauffer1999adaptive}
\bibfield{author}{\bibinfo{person}{Chris Stauffer} {and}
  \bibinfo{person}{W~Eric~L Grimson}.} \bibinfo{year}{1999}\natexlab{}.
\newblock \showarticletitle{Adaptive background mixture models for real-time
  tracking}. In \bibinfo{booktitle}{\emph{Proceedings of Computer Vision and
  Pattern Recognition}}. \bibinfo{publisher}{IEEE}, \bibinfo{address}{Fort
  Collins, CO, USA}, \bibinfo{pages}{246--252}.
\newblock


\bibitem[\protect\citeauthoryear{Sun, Tian, and Mei}{Sun et~al\mbox{.}}{2015}]%
        {sun2015anomaly}
\bibfield{author}{\bibinfo{person}{Teng Sun}, \bibinfo{person}{Hui Tian}, {and}
  \bibinfo{person}{Xuan Mei}.} \bibinfo{year}{2015}\natexlab{}.
\newblock \showarticletitle{Anomaly detection and localization by diffusion
  wavelet-based analysis on traffic matrix}.
\newblock \bibinfo{journal}{\emph{Computer Science and Information Systems}}
  \bibinfo{volume}{12}, \bibinfo{number}{4} (\bibinfo{year}{2015}),
  \bibinfo{pages}{1361--1374}.
\newblock


\bibitem[\protect\citeauthoryear{Szmit and Szmit}{Szmit and Szmit}{2012}]%
        {szmit2012usage}
\bibfield{author}{\bibinfo{person}{Maciej Szmit} {and} \bibinfo{person}{Anna
  Szmit}.} \bibinfo{year}{2012}\natexlab{}.
\newblock \showarticletitle{Usage of modified Holt-Winters method in the
  anomaly detection of network traffic: Case studies}.
\newblock \bibinfo{journal}{\emph{Journal of Computer Networks and
  Communications}}  \bibinfo{volume}{2012} (\bibinfo{year}{2012}).
\newblock


\bibitem[\protect\citeauthoryear{Uthayakumar, Vengattaraman, and
  Amudhavel}{Uthayakumar et~al\mbox{.}}{2017}]%
        {uthayakumar2017simple}
\bibfield{author}{\bibinfo{person}{J Uthayakumar}, \bibinfo{person}{T
  Vengattaraman}, {and} \bibinfo{person}{J Amudhavel}.}
  \bibinfo{year}{2017}\natexlab{}.
\newblock \showarticletitle{A simple data compression algorithm for anomaly
  detection in Wireless Sensor Networks}.
\newblock \bibinfo{journal}{\emph{International Journal of Pure and Applied
  Mathematics}} \bibinfo{volume}{117}, \bibinfo{number}{19}
  (\bibinfo{year}{2017}), \bibinfo{pages}{403--410}.
\newblock


\bibitem[\protect\citeauthoryear{Valenzise, Gerosa, Tagliasacchi, Antonacci,
  and Sarti}{Valenzise et~al\mbox{.}}{2007}]%
        {valenzise2007scream}
\bibfield{author}{\bibinfo{person}{Giuseppe Valenzise}, \bibinfo{person}{Luigi
  Gerosa}, \bibinfo{person}{Marco Tagliasacchi}, \bibinfo{person}{Fabio
  Antonacci}, {and} \bibinfo{person}{Augusto Sarti}.}
  \bibinfo{year}{2007}\natexlab{}.
\newblock \showarticletitle{Scream and gunshot detection and localization for
  audio-surveillance systems}. In \bibinfo{booktitle}{\emph{Conference on
  Advanced Video and Signal Based Surveillance}}. \bibinfo{publisher}{IEEE},
  \bibinfo{address}{London, UK}, \bibinfo{pages}{21--26}.
\newblock


\bibitem[\protect\citeauthoryear{Vitter}{Vitter}{1987}]%
        {vitter1987design}
\bibfield{author}{\bibinfo{person}{Jeffrey~Scott Vitter}.}
  \bibinfo{year}{1987}\natexlab{}.
\newblock \showarticletitle{Design and analysis of dynamic Huffman codes}.
\newblock \bibinfo{journal}{\emph{J. ACM}} \bibinfo{volume}{34},
  \bibinfo{number}{4} (\bibinfo{year}{1987}), \bibinfo{pages}{825--845}.
\newblock


\bibitem[\protect\citeauthoryear{Wang and Ahn}{Wang and Ahn}{2020}]%
        {wang2020real}
\bibfield{author}{\bibinfo{person}{Xinlin Wang} {and}
  \bibinfo{person}{Sung-Hoon Ahn}.} \bibinfo{year}{2020}\natexlab{}.
\newblock \showarticletitle{Real-time prediction and anomaly detection of
  electrical load in a residential community}.
\newblock \bibinfo{journal}{\emph{Applied Energy}}  \bibinfo{volume}{259}
  (\bibinfo{year}{2020}), \bibinfo{pages}{114145}.
\newblock


\bibitem[\protect\citeauthoryear{Widmer and Kubat}{Widmer and Kubat}{1996}]%
        {widmer1996learning}
\bibfield{author}{\bibinfo{person}{Gerhard Widmer} {and}
  \bibinfo{person}{Miroslav Kubat}.} \bibinfo{year}{1996}\natexlab{}.
\newblock \showarticletitle{Learning in the presence of concept drift and
  hidden contexts}.
\newblock \bibinfo{journal}{\emph{Machine learning}} \bibinfo{volume}{23},
  \bibinfo{number}{1} (\bibinfo{year}{1996}), \bibinfo{pages}{69--101}.
\newblock


\bibitem[\protect\citeauthoryear{Wu, Gianvecchio, Xie, and Wang}{Wu
  et~al\mbox{.}}{2010}]%
        {wu2010mimimorphism}
\bibfield{author}{\bibinfo{person}{Zhenyu Wu}, \bibinfo{person}{Steven
  Gianvecchio}, \bibinfo{person}{Mengjun Xie}, {and} \bibinfo{person}{Haining
  Wang}.} \bibinfo{year}{2010}\natexlab{}.
\newblock \showarticletitle{Mimimorphism: A new approach to binary code
  obfuscation}. In \bibinfo{booktitle}{\emph{Proceedings of the 17th conference
  on Computer and communications security}}. \bibinfo{publisher}{ACM},
  \bibinfo{address}{Chicago Illinois, USA}, \bibinfo{pages}{536--546}.
\newblock


\bibitem[\protect\citeauthoryear{Zang and Klette}{Zang and Klette}{2006}]%
        {zang2006parameter}
\bibfield{author}{\bibinfo{person}{Qi Zang} {and} \bibinfo{person}{Reinhard
  Klette}.} \bibinfo{year}{2006}\natexlab{}.
\newblock \bibinfo{booktitle}{\emph{Parameter analysis for mixture of gaussians
  model}}.
\newblock \bibinfo{type}{{T}echnical {R}eport}. \bibinfo{institution}{CITR, The
  University of Auckland, New Zealand}.
\newblock


\bibitem[\protect\citeauthoryear{Zhao, Deng, Shen, Liu, Lu, and Hua}{Zhao
  et~al\mbox{.}}{2017}]%
        {zhao2017spatio}
\bibfield{author}{\bibinfo{person}{Yiru Zhao}, \bibinfo{person}{Bing Deng},
  \bibinfo{person}{Chen Shen}, \bibinfo{person}{Yao Liu},
  \bibinfo{person}{Hongtao Lu}, {and} \bibinfo{person}{Xian-Sheng Hua}.}
  \bibinfo{year}{2017}\natexlab{}.
\newblock \showarticletitle{Spatio-temporal autoencoder for video anomaly
  detection}. In \bibinfo{booktitle}{\emph{Proceedings of the 25th
  international conference on Multimedia}}. \bibinfo{publisher}{ACM},
  \bibinfo{address}{Mountain View California, USA},
  \bibinfo{pages}{1933--1941}.
\newblock


\end{thebibliography}

\end{document}